\documentclass[aps,prd,twocolumn,reprint,amsmath,amssymb,nopacs,nofootinbib,
floatfix,eqsecnum]{revtex4-1}
\pdfoutput=1

\usepackage{mathbbol}
\usepackage[pdftex]{graphicx}
\usepackage{hyperref}
\usepackage{slashed}
\usepackage{verbatim}
\usepackage{bm}

\usepackage{amssymb}
\usepackage{amsmath}
\newcommand{\be}{\begin{equation}}
\newcommand{\ee}{\end{equation}}              
\newcommand\1{\mathbb 1}
\newcommand\0{\mathbb 0}

\def\eq#1{(\ref{#1})}
\def\eqref#1{Eq.~(\ref{#1})}

\def\fig#1{Fig.~\ref{#1}}
\def\sec#1{Sec.~\ref{#1}}

\def\pa{\partial}
\def\b{\beta}
\def\T{T} % temperature
\def\x{\vec{x}}

\def\t{t} %time

\def\l{\lambda} % eigenvalues
\def\ps{\psi} % eigenfunctions
\def\psii{\check{\psi}}

\def\ch{\rho} % pseudo-density

\def\ds{\slashed D} % Dirac operators
\def\dsmu{{\slashed D}(\mu)}

\def\p{\varphi} % boundary angle
\def\bp{\bar\varphi}

\def\ph{\chi} % left eigenmode

\def\P{P_\infty} % Polyakov loop
\def\bP{\bar{P}_\infty}

\def\al{\alpha}

\def\one{\Eins}

\usepackage{color}
\definecolor{mygreen}{rgb}{0.1,0.4,0}

\newcommand{\tr}{\text{tr}}
\newcommand{\m}{\mu_\p}

\begin{document}

\title{Topological zero modes at nonzero chemical potential}
\author{Falk Bruckmann}
\author{Rudolf R\"odl}
\author{Tin Sulejmanpasic}
\affiliation{Institute for Theoretical Physics, University of
  Regensburg, 93040 Regensburg, Germany}
%\date{\today}

\begin{abstract}

We give analytical and numerical solutions for the zero modes of the Dirac operator in topological SU(2) gauge backgrounds at nonzero chemical potential. Continuation from imaginary to real chemical potential is 
used to systematically derive analytical zero modes for calorons at arbitrary holonomy and, in particular limits, for instantons and dyons (magnetic monopoles). For the latter a spherical ansatz is explored as well. All the zero mode exhibit stronger peaks at the core and negative regions in their densities. We discuss the structure of the corresponding overlap matrix elements. For discretized calorons on the lattice we consider the staggered operator and show that it does possess (quartet quasi-)zero modes, whose eigenmodes agree very well with the continuum profiles, also for nonzero chemical potential. 

\end{abstract}

\maketitle

\section{Introduction}

The thermodynamics of Quantum Chromodynamics (QCD) has attracted a lot of attention in recent years, both experimentally and theoretically. While at high temperature the transition to a phase of deconfined gluons and quarks obeying chiral symmetry is rather well understood, the exploration of a finite quark density is still immature. Theoretical progress in this part of the QCD phase diagram has mainly been hindered by the failure of the most powerful nonperturbative method from first principals, lattice gauge theory. At finite chemical potential the Dirac operator is not purely antihermitian %(not $\gamma_5$-hermitian) 
and its complex determinant prevents an interpretation as a probability weight. This is the famous sign-problem. Ways to proceed include reweighting and expansion techniques on the lattice, complex Langevin dynamics, and studies of finite density in diagrammatic and related approaches and approximations to QCD or similar theories such as random matrix theory, for reviews see e.g.~\cite{deForcrand:2010ys,Fukushima:2010bq}. 

Here we investigate the semiclassical approach to QCD, which is based on solutions of the Euclidean Yang-Mills equations. At zero temperature these are the celebrated instantons \cite{Belavin:1975fg}, localized lumps of action density characterized by an integer topological charge. At finite temperature, the corresponding solutions are called calorons \cite{Harrington:1978ve}. Calorons have been found to be governed by an additional parameter, the holonomy, and consist of constituent dyons (magnetic monopoles, see below) \cite{Kraan:1998pm,Lee:1998bb}. Individual dyons constitute static solutions at finite temperature. All these configurations are selfdual or antiselfdual and thus minima of the Yang-Mills action in the given topological sector. 

The semiclassical theory of instantons and calorons with trivial holonomy has been studied at zero and finite temperature  \cite{Schafer:1993ra,Schafer:1994nv,Schafer:1994nw,Schafer:1995pz,Schafer:1995uz} as well as nonzero chemical potential \cite{Rapp:1997zu,Schafer:1997ep,Schafer:1998up,Alford:1997zt,Carter:1998jw} (see \cite{Schafer:1996wv} for a review). The corresponding mechanism of chiral symmetry breaking is rather robust: exact fermionic zero modes become low-lying modes and thus generate a nonzero density of states at zero Dirac eigenvalue, which via the Banks-Casher relation is linked to the chiral condensate. Confinement, on the other hand, could not be obtained from instantons nor calorons of trivial holonomy. This changes drastically for calorons at non-trivial holonomy, as has first been shown semi-analytically in \cite{Gerhold:2006sk}. The lesson is that the holonomy plays an important role with the confinement mechanism becoming close to the old Dual superconductor picture. Promising results in this context are the possibility to show confinement analytically from the moduli space metric of dyons \cite{Diakonov:2007nv} and even from a non-interacting gas of dyons \cite{Bruckmann:2011yd}. Although the interactions between calorons or dyons are not fully worked out, there are more hints on confinement, namely from the quantum weight \cite{Diakonov:2004jn}, the detection of calorons in lattice quantum configurations \cite{Ilgenfritz:2002qs,Ilgenfritz:2004zz,Ilgenfritz:2006ju} and the vortex content of calorons \cite{Bruckmann:2009pa}. In the high temperature phase either dyons of particular kinds are suppressed \cite{Bornyakov:2008im,Bruckmann:2009ne} or dyons form molecules due to their fermionic interactions \cite{Shuryak:2012aa}. 

Instantons, calorons and dyons also share the support for fermionic zero modes, which are present due to the index theorems \cite{Atiyah:1970ws,Callias:1977kg,Nye:2000eg,Poppitz:2008hr}. These modes localize to the background as can be seen from the existing explicit solutions \cite{Jackiw:1975fn,GarciaPerez:1999ux,Bruckmann:2003ag}. Generalizations of the index theorem to nonzero chemical potentials have been discussed in \cite{Gavai:2009vb,Kanazawa:2011tt}. Our results confirm these as far as instantons/calorons of unit charge and dyons of fractional charge are concerned. For the semiclassical approach at finite density it is necessary to know the zero eigenmodes, both for the formation of the condensate and the fermionic interaction of the objects.

\begin{figure*}[t]\centering\vspace*{-0.2cm}
   \includegraphics[width=2\columnwidth]{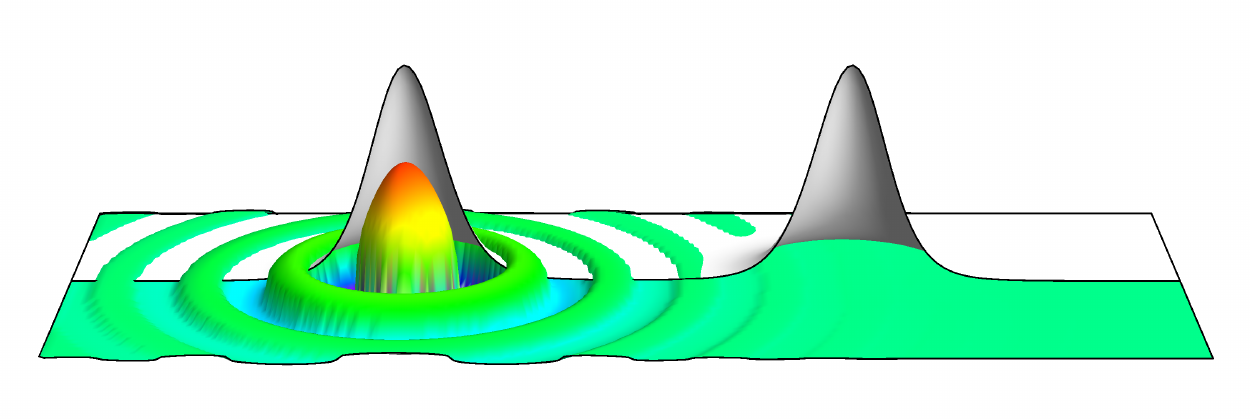} \vspace*{-0.5cm}
   \caption{Zero mode in a background of a caloron with maximally nontrivial holonomy and $\mu=4T$ (colored, magnitude rescaled to its third root) together with the caloron action density revealing two constituent dyons (grey, cut along the caloron axis).}
   \label{fig:appetizer}
\end{figure*}

In this paper we present analytical and numerical results for SU(2) zero modes at nonzero chemical potential. We show that the analytic continuation from imaginary chemical potential provides exactly the particular (bi-) ortho\-nor\-ma\-li\-za\-tion necessary for the eigenmodes at real chemical potential. Together with the results from the ADHM-Nahm formalism \cite{Atiyah:1978ri,Nahm:1979yw} we systematically obtain the zero modes in caloron backgrounds at arbitrary holonomy and chemical potential. In particular limits we confirm the zero modes of trivial holonomy calorons and instantons known in the literature \cite{AragaodeCarvalho:1980de,Abrikosov:1983rh,Cristoforetti:2011fq} as well as for dyons, in which case we show that the spherically symmetric ansatz \cite{Jackiw:1975fn,Shuryak:2012aa} works. The typical features of these modes are stronger peaks at and oscillations away from the core, with the period of oscillation $\propto 1/\mu$, cf.~\fig{fig:appetizer}. This will have a wash-out effect in the fermionic interactions of these objects, which we briefly discuss. 

We also present numerical results for calorons discretized on the lattice utilizing the staggered Dirac operator. Although the latter has four tastes and no exact zero modes of definite chirality, we identify modes smaller than the rest of the spectrum by several orders of magnitude. These quasi-zero modes  persist at nonzero chemical potential, too. Taking the taste quartets together, the profiles agree very well with the profiles of the continuum zero modes. 

This paper is organized as follows: In the next section we discuss properties of the Dirac operator and its eigenmodes at nonzero chemical potential. After that in Sec.~\ref{sec_continuation} we perform the analytic continuation from imaginary to real chemical potential. Sec.~\ref{sec_analytic} contains the corresponding analytic results for zero modes in dyon, caloron and instanton backgrounds. Sec.~\ref{sec_numerics} is devoted to the lattice results for the staggered eigenmodes. Sec.~\ref{sec_overlap} gives first results for the overlap matrix elements of dyon zero modes. We summarize our findings in Sec.~\ref{sec_summary}. 

\section{Dirac operator with chemical potential} 

In the continuum Dirac operator the chemical potential is added to the covariant derivative 
%[convention of Gattringer/Lang]:
\begin{align}
 \dsmu=\gamma_\nu D_\nu-\gamma_0\mu\,,\quad 
 D_\nu=\pa_\nu+i A_\nu
\label{eq_dirac_equation} 
\end{align}
(for our lattice Dirac operator see \sec{subsec_staggered_def}).
We will always focus on the massless case and analyse eigenmodes
\begin{align}\label{eq_eigenmode_right}
 \dsmu\ps_n(x;\mu)=\l_n(\mu)\ps_n(x;\mu)\,,
\end{align}
in particular zero modes, $\l_0=0$ (arguments will be neglected whenever convenient). 

At finite temperature $T=1/\b$, physical modes are antiperiodic in the temporal direction
\begin{align}\label{eq_period_first}
 \ps(\t+\b,\x)=-\ps(\t,\x)\,,
\end{align}
but we will also consider modes periodic up to a phase. 

\subsection{Symmetries}

Since the Dirac operator is chiral at arbitrary chemical potential, $\{\dsmu,\gamma_5\}=0$, its eigenvalues always come in pairs $\pm\l$.

While the Dirac operator at vanishing $\mu$ is antihermitian with its eigenvalues purely imaginary, this important property is lost for real $\mu$. However, for the hermitian conjugate operator the following relation holds,
\begin{align}\label{eq_related_operator}
 \ds^\dagger(\mu)=-\ds(-\mu)\,,\quad 
 \l^*(\mu)=-\l(-\mu)\,,\quad
 (\mu\in \mathbb{R})
\end{align}
(for complex $\mu$'s replace $-\mu$ by $-\mu^*$).

For gauge group SU(2), due to the pseudoreal nature of this group, the Dirac operator with (real) chemical potential obeys an anti-unitary symmetry
\begin{align}\label{anti_symm_cont}
\ds^*(\mu)=-W^\dagger\dsmu W\qquad (SU(2),\:\mu\in\mathbb{R})
\end{align}
where $W$ is unitary and $\mu$-independent\footnote{In the Weyl representation one has  $W=\gamma_2\gamma_0\otimes\tau_2$, with the charge conjugation $\gamma_2\gamma_0$ relating gamma matrices to their complex conjugates and the second Pauli matrix $\tau_2$ relating the Pauli matrices as generators in color space to their complex conjugates.} with $W^* W=1$. 
In Random matrix theory this symmetry qualifies $i\ds(0)$ as belonging to the universality class of the Gaussian orthogonal ensemble. 
Moreover, in a particular basis $i\ds(0)$ is real symmetric with real eigenmodes (see below; for the different anti-unitary symmetry of our lattice Dirac operator see \sec{subsec_staggered_def}). In the next subsection, we will use this symmetry to show that the SU(2) zero modes have real densities.

For operators without definite hermiticity one has to distinguish eigenvalues with corresponding left and right eigenmodes from singular values as will be discussed in the next subsections.

\subsection{Left and right eigenmodes}
\label{sec:laft_and_right}

The right eigenmodes of the Dirac operator are given through \eqref{eq_eigenmode_right}, whereas the left eigenmodes are defined analogously
\begin{align}\label{eq_eigenmode_left}
 \ph_m^\dagger(\mu)\dsmu=\ph_m^\dagger(\mu)\l_m(\mu)\,.
\end{align}
For our specific case, the left eigenmodes are related to the right eigenmodes at opposite $\mu$
\begin{align}
 \ph_m(\mu)= \ps_m(-\mu)\;,\quad(\mu\in\mathbb{R})
\end{align}
(which follows from the hermitian conjugate of \eqref{eq_eigenmode_right} and the property \eq{eq_related_operator}). Thus, one may replace one by the other.

From the difference of the scalars product of $\ph_m^\dagger(\mu)$ with \eq{eq_eigenmode_right} and of \eq{eq_eigenmode_left} with $\ps_n(\mu)$ one immediately concludes that right and left eigenvalues agree and that left and right eigenmodes are orthogonal. 
Defining the (pseudo-)density 
\begin{align}\label{eq_left_vs_opposite_mu}
 \ch_{mn}(x;\mu)
 &\equiv \ps_m^\dagger(x;-\mu)\ps_n(x;\mu)\nonumber\\
 &= \sum_a \ps_m^{a}(x;-\mu)^*\ps_n^a(x;\mu)
\end{align}
where the index $a$ stands for both color and spin, the $x$-integral over $\ch_{mn}$ is
\begin{align}\label{eq_chi_integral}
 \int \!\!d^4 x\, \ch_{mn}(x;\mu)=\delta_{mn}
\end{align}
Such a basis is called {\it bi-orthonormal}. At vanishing chemical potential this of course reduces to the orthonormal basis of eigenmodes of the purely antihermitian Dirac operator. 

Note that for $\mu\neq 0$ the densities $\ch_{mm}(x)$ of individual eigenvalues need not be real everywhere in space-time, although their integral is positive. 

Densities of SU(2) zero modes can be made real thanks to the symmetry \eq{anti_symm_cont}.
For this gauge group $ i\ds(\mu)$ can be transformed to $S+\mu A$ with real symmetric $S$ and real antisymmetric $A$ \cite{Bloch:2012wh}. For a real matrix the zero modes can be chosen real (as the complex conjugate of a zero mode is a zero mode again). Hence the density
$\ch_{00}(x)$ is real in this and every other basis (as this fermion bilinear is an invariant). 
How the densities come out real in an arbitrary basis is shown in App.~\ref{sec:reality}. Still, negative densities can, and will, occur.
 
\subsection{Singular value decomposition} 

Arbitrary matrices can be written in a singular value decomposition, for the Dirac operator
\begin{align}\label{eq_svd}
 U^\dagger \ds V %=\S
 =\mbox{diag}(\xi_1,\xi_2,\ldots)\,,\quad \xi_n\geq 0
\end{align}
where $U$ and $V$ are unitary transformations 
and the entries $\xi_n$ -- which are non-negative -- are called singular values. The columns $u_n$ of $U$ and $v_n$ of $V$ are called left and right singular modes (vectors), respectively. They are orthonormal among themselves, but not necessarily to each other. 

For generic configurations at nonzero $\mu$, these singular values and modes are not related to the eigenvalues and eigenmodes of the operator $\ds$ itself, rather to those of its product with $\ds^\dagger$: the left and right singular modes are the eigenmodes of the hermitian operators $\ds\ds^\dagger$ and $\ds^\dagger\ds$, respectively, and the nonzero singular values $\xi_n$ are the square roots of the eigenvalues of these operators (which coincide between $\ds\ds^\dagger$ and $\ds^\dagger\ds$).

Zero singular and eigenvalues, however, constitute an exception as the zero modes are identical, $v_0=\ps_0$ and $u_0=\ph_0$
\footnote{That a right zero eigenmode of $\ds$ is also a right zero eigenmode of $\ds^\dagger\ds$ and thus a zero singular mode follows immediately; the opposite direction of the proof uses that $\ds^\dagger\ds\ps=0$ induces $|\ds\ps|^2=0$, which means that $\ds\ps$ must vanish. The same argument applies to left zero eigenmodes.}. 
Thus, in order to determine zero modes of the Dirac operator, both the eigenvalue and the singular value method can be used.   

At the fundamental level, the singular value decomposition is slightly more powerful, as it always exists and its complete and orthonormal basis can be used to decompose the fermion fields in the path integral. Eigenmodes, on the other hand, do not always provide a complete basis. The simplest example for a matrix of this kind is $\left(\begin{matrix}0&1\\0&0\end{matrix}\right)$, 
a $4\times 4$ example is discussed in \cite{Kanazawa:2011tt}. Finite $\mu$ Dirac operators with this property occur in principle, but the corresponding configurations are non-generic. 

\subsection{Index theorem}
  
We start by recalling the case of vanishing $\mu$. In a chiral basis the Dirac operator can be written as 
\begin{align}
 \ds=\left(\begin{array}{cc} 0 & D_L \\ D_R & 0\end{array}\right) 
\end{align}
Then $\gamma_5=\mbox{diag}(1_2,-1_2)$ and left and right handed zero modes consist only of lower and upper components, respectively. Let us denote their numbers by $n_{L,R}=\mbox{dim}\,\mbox{ker}\,D_{L,R}$ and specialize to positive topological charge $Q$ for simplicity. 

The index theorem reads $n_R-n_L=Q$. It is fulfilled by generic configurations through possessing $Q$ right handed, but no left handed zero modes.

With nonzero chemical potential the index theorem has been discussed in Refs.~\cite{Gavai:2009vb,Kanazawa:2011tt}. It now also incorporates the hermitean conjugates of $D_{L,R}$, for which the relation \eq{eq_related_operator} implies
\begin{align}\label{eq_hermconj_LR}
 D_L^\dagger(\mu)=-D_R(-\mu)\,,\quad D_R^\dagger(\mu)=-D_L(-\mu)
\end{align}
A straightforward generalization of the index theorem to real chemical potential has been obtained in \cite{Gavai:2009vb}
\begin{align}
 n_R(\mu)-n_L(\mu)=Q\,,\quad n_R(-\mu)-n_L(-\mu)=Q
\end{align}
where we have trivially added the index theorem at opposite $\mu$. The more appropriate index theorem claimed in \cite{Kanazawa:2011tt} (Eq.~(D.3)) -- derived via the singular value decomposition -- translates via \eq{eq_hermconj_LR} into
 \begin{align}
 n_R(\mu)-n_L(-\mu)=Q\,,\quad n_R(-\mu)-n_L(\mu)=Q
\end{align}
The two index theorems agree for almost all configurations \cite{Kanazawa:2011tt}. 

Our background configurations obey both index theorems in the generic way by possessing one right handed, but no left handed zero mode. Due to the latter, they are not sensitive to the difference of the two index theorems. 

\section{Analytic continuation in the chemical potential}
\label{sec_continuation}

In this section we will demonstrate that eigenmodes at imaginary chemical potential can be analytically continued to yield eigenmodes at real chemical potential obeying the correct bi-orthonormalization. 

Let us consider the operator
\begin{align}\label{eq_continue_operator_first}
 \ds'(\bp)\equiv\gamma_\nu D_\nu+i\gamma_0\bp\T=\ds(-i\bp\T)
\end{align}
which is nothing but the Dirac operator with an imaginary chemical potential $-i\bp\T$. The minus sign in the argument on the right is for later convenience. At real $\bp$ the operator $\ds'$ is purely anti-hermitean, thus in its eigenequation
the eigenvalues are purely imaginary and the eigenmodes are orthonormal in the conventional sense, i.e. the densities
\begin{align}
 \ch'_{mn}(x;\bp)=\sum_a \ps_m^{\prime a}(x;\bp)^*\ps_n^{\prime a}(x;\bp)
\end{align}
are non-negative at every space-time point $x$ and integrate to
\begin{align}\label{eq_chi_integral_imag}
 \int \!\!d^4 x\, \ch'_{mn}(x;\bp)=\delta_{mn}\,.
\end{align}

Now we analytically continue $\bp$ to imaginary values $i\mu/\T$ assuming this is possible continuously. The Dirac operator becomes that at real chemical potential,
\begin{align}
 \ds'(i\mu/\T)=\ds(\mu)
\end{align}
cf.\ \eqref{eq_continue_operator_first}. Then, analytically continuing the eigenequation, the $\ps'_n(x;i\mu/\T)$ obey the eigenequation of $\ds(\mu)$ and are thus candidates for eigenmodes,
\begin{align}\label{eq_psi_continue_psi}
 \ps'_n(x;i\mu/\T)=\ps_n(x;\mu)\,.
\end{align}
That these modes are indeed eigenmodes, all that needs to be checked is their bi-orthonormalization. For that we carefully perform the analytic continuation in the densities
\begin{align}\label{eq_cont_details}
 \ch'_{mn}( i\mu/\T)
  &=\sum_a \ps_m^{\prime a*}(\bp)\big|_{\bp=i\mu/\T}\ps_n^{\prime a}(\bp)\big|_{\bp= i\mu/\T}\nonumber\\
  &=\sum_a \ps_m^{\prime a}(-i\mu/\T)^*\ps_n^{\prime a}(i\mu/\T)\nonumber\\
  &=\sum_a \ps_m^{a}(-\mu)^*\ps_n^{a}(\mu)=\ch_{mn}(\mu)
\end{align}
valid for all $x$ (suppressed).
The opposite $\mu$-arguments of the two factors\footnote{In going from the first to the second line in \protect\eq{eq_cont_details} we have used that $\ps^{\prime}$ is a smooth function in real $\bp$ with Taylor expansion $\ps^{\prime}(\bp)=\sum_n c_n\bp^n$ and its complex conjugate reads $\ps^{\prime *}(\bp)=\sum_n c_n^*\bp^n$. For the analytic continuation of the latter we plug in an imaginary $\bp$, $\ps^{\prime *}(\bp)|_{\bp=i\mu/\T}=\sum_n c_n^*(i\mu/\T)^n$, which agrees with the complex conjugate of $\ps^{\prime}(-i\mu/\T)$.} are exactly what is needed to match the pseudo-density $\ch_{mn}(\mu)$ at real chemical potential, cf.\ \eqref{eq_left_vs_opposite_mu}. Then the desired bi-orthonormalization of it,  \eqref{eq_chi_integral}, is inherited from the analytic continuation of the conventional orthonormalization \eq{eq_chi_integral_imag} of eigenmodes of the purely antihermitian operator $\ds'$. Moreover, we have started with eigenmodes $\ps'$ being antiperiodic in $x_0$ and so are their analytic continuations; actually all kinds of boundary conditons are continued to the eigenmodes at real chemical potential. Hence, {\it the modes $\ps'_n(\bar\p)$ analytically continued to  $\bar\p=i\mu/\T$ are indeed the eigenmodes at real chemical potential}, i.e.\ \eqref{eq_psi_continue_psi} does hold. This continuation applies to all spin and color components of the eigenmodes and, of course, also to the eigenvalues, i.e.\  $\l'_n(i\mu/\T)=\l_n(\mu)$.

The dimensionless parameter $\bar\p$ in \eqref{eq_continue_operator_first} can be interpreted as a constant trace part of the gauge field and can also be transfered to the temporal boundary condition defining
\begin{align}\label{eq_period_trafo}
 \psii_n(x;\bar\p)= e^{ i\bar\p \t/\b}\ps'_n(x;\bar\p)
\end{align}
which are eigenmodes of the Dirac operator at vanishing chemical potential, but periodic up to a phase $\bar\p$ (on top of the antiperiodicity)
\begin{align}\label{eq_period_second}
 \psii_n(\t+\b,\x;\bar\p)
  &=-\,e^{i\bar\p}\psii_n(\t,\x;\bar\p)\nonumber\\
  &=e^{-i\p}\psii_n(\t,\x;\bar\p)\,,\quad\bp=\pi-\p
\end{align}
where we have also defined the total phase $\p$ for later use.
The pseudo-densities of these mode still agree with those of the original eigenmodes, $\check{\ch}=\ch'$.

In particular, knowing the zero modes in an analytic fashion for (small) imaginary chemical potentials, one immediately gets the zero modes of the desired Dirac operator at real chemical potential. In \sec{sec_profiles_caloron} we will make use of this analytic continuation, since in caloron backgrounds (and limits thereof) the zero modes are known analytically for whole intervals of imaginary chemical potentials/boundary conditions.

\section{Analysis of zero modes in specific topological backgrounds}
\label{sec_analytic}

\subsection{Single dyon zero mode}\label{sec:dyons}

Here we discuss the solution of the zero mode on a single dyon for arbitrary boundary condition and with arbitrary chemical potential $\mu$. By dyon we mean a magnetic monopole obeying the BPS condition, which in a 3+1-dimensional language amounts to possessing an (Euclidean) electric charge of same magnitude\footnote{It was argued in \cite{Poppitz:2012sw} that the term ``dyon'' is not entirely appropriate for these objects. They suggest that the term ``monopole-instanton'' is more accurate. Nevertheless we will use the term dyon in this work to mean the constituent of a finite temperature instanton, see below.}.
The derivation of the zero mode makes use of the spherical symmetry and closely follows \cite{Jackiw:1975fn,Shuryak:2012aa}.

We work in the BPS limit and in a hedgehog gauge. The (anti)self-duality equations for the gauge fields are solved for by (see e.g. \cite{Diakonov:2009jq})
\begin{subequations}\label{eq:radialgaugefields_both}
\begin{align}
 \!\!&A_i^a=\mathcal{A}\epsilon_{aik}\hat{ r}^k\,, &&\!\!\! \mathcal{A}=\frac{1}{r}\left(1-\frac{vr}{\sinh(vr)}\right)%\;,\\
 \stackrel{r\to \infty}{\longrightarrow}\frac{1}{r}\,,\\
 \!\!&A_0^a=\mathcal{H}\hat{ r}^a\,,  &&\!\!\!
 \mathcal{H}=\mp\frac{1-vr\coth(vr)}{r}%\;,
 \stackrel{r\to \infty}{\longrightarrow}\pm v\,,\label{eq:radialgaugefields}
\end{align}
\end{subequations}
with $\hat r=\vec{x}/r$ being the unit vector in the radial direction and $A_\mu=A^{a}_\mu\frac{\tau^a}{2}$ with $\tau^a$ the Pauli matrices. The upper sign refers to the selfdual and the lower to the anti-selfdual field equations\footnote{We define selfduality as $F_{\mu\nu}=\epsilon_{\mu\nu\rho\sigma}F_{\nu\sigma}/2$, where $\mu,\nu,\rho,\sigma=0,\dots,3$. From $\epsilon_{0123}=1$ this yields, e.g. $F_{01}=F_{23}$. Note that self-duality equations with $\mu,\nu,\rho,\sigma=1,\dots,4$ and $\epsilon_{1234}=1$ yield $F_{41}=-F_{23}$.}, and $v$ is the asymptotic value of $A_0$,
\be\label{eq:higgs}
v=\lim_{r\rightarrow\infty} \sqrt{(A_0^a)^2}\;.
\ee
playing the role of a ``Higgs'' vacuum expectation value. For simplicity we will take $v>0$. The vev  also enters the action and topological charge of the dyon, which we define in the four-dimensional language:
\begin{align}
 \!\!\!\!S=\frac{1}{2}\int_0^\b \!\!dt \!\int\!\! d^3x\,\tr F_{\mu\nu}^2=4\pi\b v\,,\:Q=\frac{S}{8\pi^2}=\frac{v}{2\pi T}
\label{eq:dyon_action_topcharge}
\end{align}
We restrict ourselves to $v\le2\pi T$, such that $Q\in[0,1]$.

For the zero modes we make the usual radial ansatz for Dirac spinors (see \cite{Jackiw:1975fn,Shuryak:2012aa,Shnir:2005xx})
\be\label{eq:ansatz}
\psii^{A}_{\alpha}=\Big[\Big(\alpha_1(r)\,\one+\alpha_2(r)\,{\hat r}\cdot\vec{\tau}\Big)\epsilon\Big]_{A\alpha}e^{-i\p t/\beta}\;,
\ee
where $A$ is the color index and $\alpha$ the spin index, and $\epsilon=i\tau_2$ is a completely antisymmetric rank-2 tensor. Obviously these functions are periodic up to a phase $e^{-i\p}$, cf.\ (\ref{eq_period_second}). 
We stress that the gauge background is a -- so far -- $t$-independent embedding of a three-dimensional dyon into $S^1\times R^3$. 
 
The Dirac equation (\ref{eq_dirac_equation}) for left and right-handed spinors is
\begin{subequations}
\begin{align}
((\sigma^\mu)_{\alpha\beta}(D_\mu)_{AB}-\mu\delta_{AB}\delta_{\alpha\beta})(\psi_R)^{B}_\beta&=0\;,\\
((\bar\sigma^\mu)_{\alpha\beta}(D_\mu)_{AB}-\mu\delta_{AB}\delta_{\alpha\beta})(\psi_L)^{B}_\beta&=0\;.
\end{align}
\end{subequations}
Here $\sigma^\mu=(\one,-i\tau^i),\, \bar\sigma^\mu=(\one,i\tau^i)$ and we get the following equations for the functions $\alpha_{1,2}(r)$
\begin{subequations}\label{eq:Dirac1}
\begin{align}
\!\!\!&\frac{d\alpha_1(r)}{dr}+\frac{\pm\mathcal H+2\mathcal A}{2}\alpha_1-i(\mu+i\p T)\alpha_2=0\;,\\
\!\!\!&\frac{d\alpha_2(r)}{dr}+\left(\frac{\pm\mathcal H-2\mathcal A}{2} +\frac{2}{r}\right)\alpha_2-i(\mu+i\p T)\alpha_1=0\;.
\end{align}
\end{subequations}
where the upper sign is to be taken for the left, and the lower for the right handed spinor component. One can see once more that the boundary phase $\p$ is equivalent to an imaginary chemical potential $i\p T$. In the following we will therefore use
\begin{equation}
 \m=\mu+i\p T
\label{eq:def_eta}
\end{equation}

To discuss the normalizability of solutions, let us specialize to $\p=0$ for the moment. While $\mathcal{A}$ decays for large distances, $\mathcal{H}$ approaches $v$ for self-dual and $-v$ for anti--self-dual dyons, respectively, see \eqref{eq:radialgaugefields}. In the asymptotic linear system of differential equations \eq{eq:Dirac1} the chemical potential gives rise to trigonometric functions. The $\mathcal{H}$-term leads to exponential functions and thus determines, whether the solutions are normalizable. For the latter the equations must read $d\alpha_.(r)/dr+v\alpha_./2+\ldots=0$. Hence a normalizable solution for the left and right spinor exists if the background field is self-dual or anti--self-dual, respectively,
and the solutions behave like $e^{-vr/2}$ asymptotically.

The full equations are solved by \cite{Shuryak:2012aa}
\begin{subequations}\label{eq:dyonsol}
\begin{align}
&\alpha_{1,2}(r;\mu)=c\,\frac{\chi_{1,2}(v r)}{\sqrt{vr \sinh(vr)}}\;,\\
&\chi_1(rv)=\Big[2\,\frac{\m}{v}\,\sin(r\m)+\tanh(vr/2)\cos(r\m)\Big]\;,\\
&\chi_2(rv)=\mp i\Big[2\,\frac{\m}{v}\,\cos(r\m)-\coth(vr/2)\sin(r\m)\Big]\;.
\end{align}
\end{subequations}
In these formulae, the boundary condition $\p$ has been reintroduced, as the imaginary part of $\m$, cf. (\ref{eq:def_eta}). It leads to exponential parts $e^{\p T r}$, which compete with the exponential decay $e^{-vr/2}$. Consequently, the solution is normalizable only if $\p\in(-v/2T,v/2T)$, i.e.\ around periodic boundary conditions. Later we will consider modifications leading to antiperiodic zero modes. 

Since we are interested in the dependence on $\mu$ we will set again $\p=0$ in what follows, i.e.\ analyze periodic zero modes.

As we discussed in \sec{sec:laft_and_right}, the relevant density of these modes is given by $\psi_0^\dagger(-\mu)\psi_0(\mu)=\ch_{00}(\mu)$,
\begin{align}
\ch_{00}(r;\mu)&=\alpha_1^*(r;-\mu)\alpha_1(r;\mu)+\alpha_2^*(r;-\mu)\alpha_2(r;\mu)\label{eq:alphas}\\
&\stackrel{\p=0}{=}|\alpha_1(r;\mu)|^2-|\alpha_2(r;\mu)|^2\nonumber\,.
\end{align}
The normalization in the sense of $\int d^4x\;\rho_{00}=1$, cf.\ \eqref{eq_chi_integral}, fixes the constant
\begin{align}\label{eq:dyonnorm}
c=\sqrt{v^3T/(4\pi)}\,.
\end{align}

\begin{figure}[t] 
   \centering
   \includegraphics[width=0.95\columnwidth]{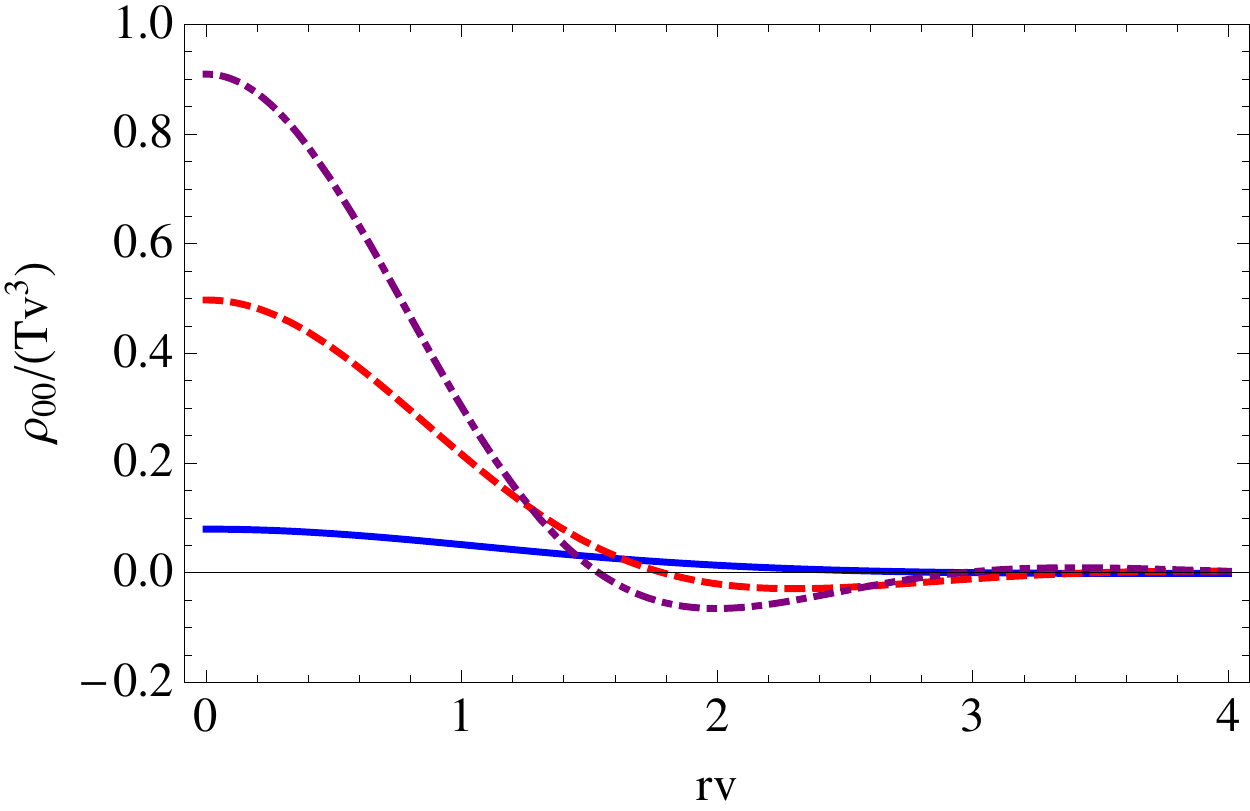} 
   \caption{Densities $\ch_{00}(r;\mu)$ of periodic dyon zero modes, \eqref{eq:dyonsol} with $\p=0$, as a function of $rv$ for $\mu=0.5v$ (solid blue), $\mu=v$ (dashed red) and $\mu=1.2v$ (dot-dashed purple).}
   \label{fig:zerodens}
\end{figure}

From \eqref{eq:alphas} one can already see the possibility for this density to become negative, through $\alpha_2$. Consistently, at vanishing $\mu$ the ($\mu$-odd) function $\alpha_2$ vanishes, and the density is positive definite. Profiles of periodic zero modes for several values of the chemical potential are plotted in \fig{fig:zerodens}.

\begin{figure}[t] 
   \centering
   \includegraphics[width=0.95\columnwidth]{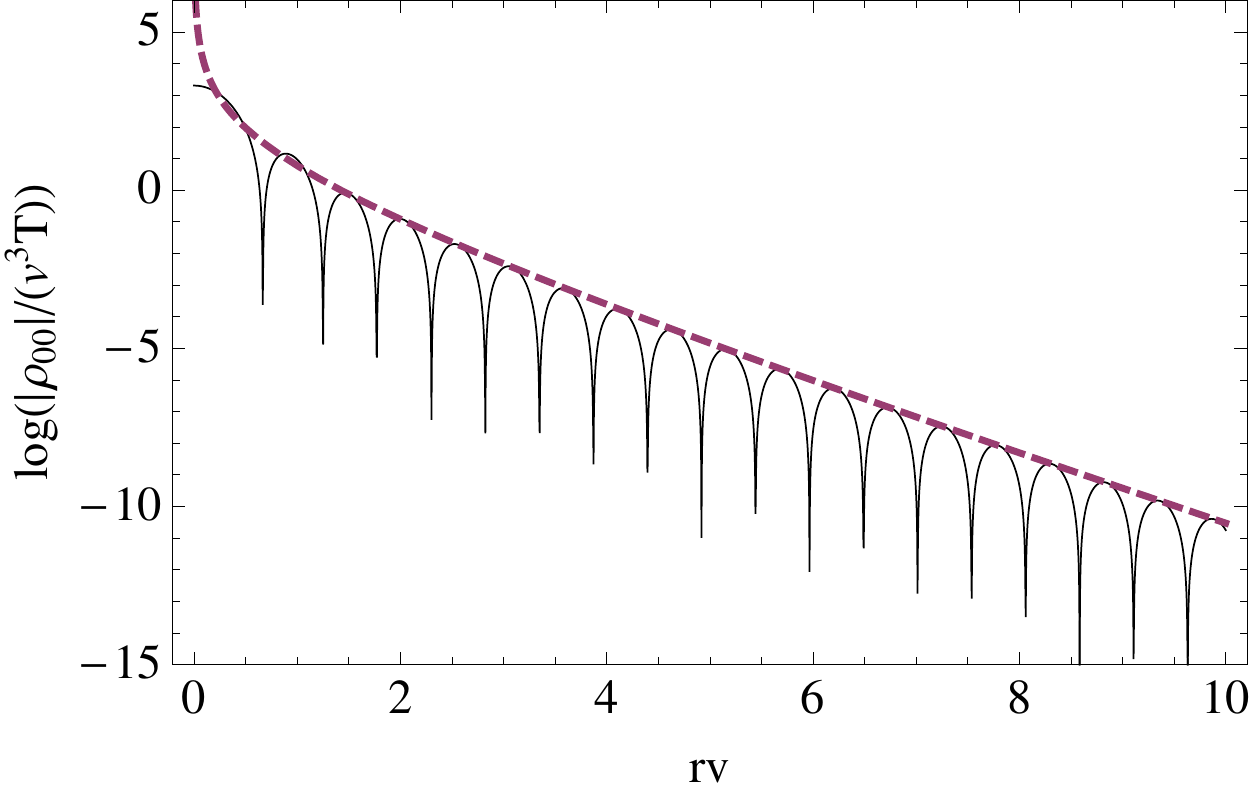} 
   \caption{Logarithmic plot of the absolute value of the density  $\rho_{00}(r;\mu)$ for $\mu=3 v$ (solid) and that of the envelope \eqref{eq:envelop} (dashed).}
   \label{fig:zerodenslog}
\end{figure}

The solutions clearly have an exponential envelope, of the asymptotic shape $e^{-vr/2}/\sqrt{r}$, see \fig{fig:zerodenslog}. For the prefactor in the envelope one has to consider $|\chi_1|^2-|\chi_2|^2$. For large $vr$ one can set $\tanh(vr/2)=1=\coth(vr/2)$ up to exponentially small corrections and get
\begin{align}\label{eq:chipart}
 |\chi_1(rv)|^2-|\chi_2(rv)|^2&=\left(1+4\,\frac{\mu^2}{v^2}\right)\sin(\delta(\mu/v)+2\mu r)\\ 
 \delta(\mu/v)&\equiv \arctan\frac{1-4\mu^2/v^2}{4\mu/v}
\end{align}
The factor in the parenthesis bounds this term and from Eqs.~(\ref{eq:dyonsol}, \ref{eq:dyonnorm}) we get
\begin{align}
\ch_{00}^{\text{env}}(r;\mu)
 =\frac{T}{2\pi}(v^2+4\mu^2)\frac{e^{-vr}}{r} 
\label{eq:envelop}
\end{align}
again up to exponential corrections in $vr$. This envelope has been included in \fig{fig:zerodenslog}.

Another property of the zero mode is that its value at the dyon core grows with the chemical potential, which can also be seen in the examples of \fig{fig:zerodens}. Taking the limit $r\to 0$ in the solution \eqref{eq:dyonsol} it is easy to see that
\begin{align}
 \ch_{00}(0;\mu)
 =\frac{T}{16\pi v}\left|v^2+4\mu_\p^2\right|^2\,.
\end{align}

Probably the most characteristic feature of the zero mode density are its negative regions. For any nonzero chemical potential there are infinitely many negative regions (recall that the far away regions have exponentially small densities though), separated by infinitely many zeroes. This is very obvious in a logarithmic plot like \fig{fig:zerodenslog}.

\begin{figure}[htbp] 
   \centering
   \includegraphics[width=0.95\columnwidth]{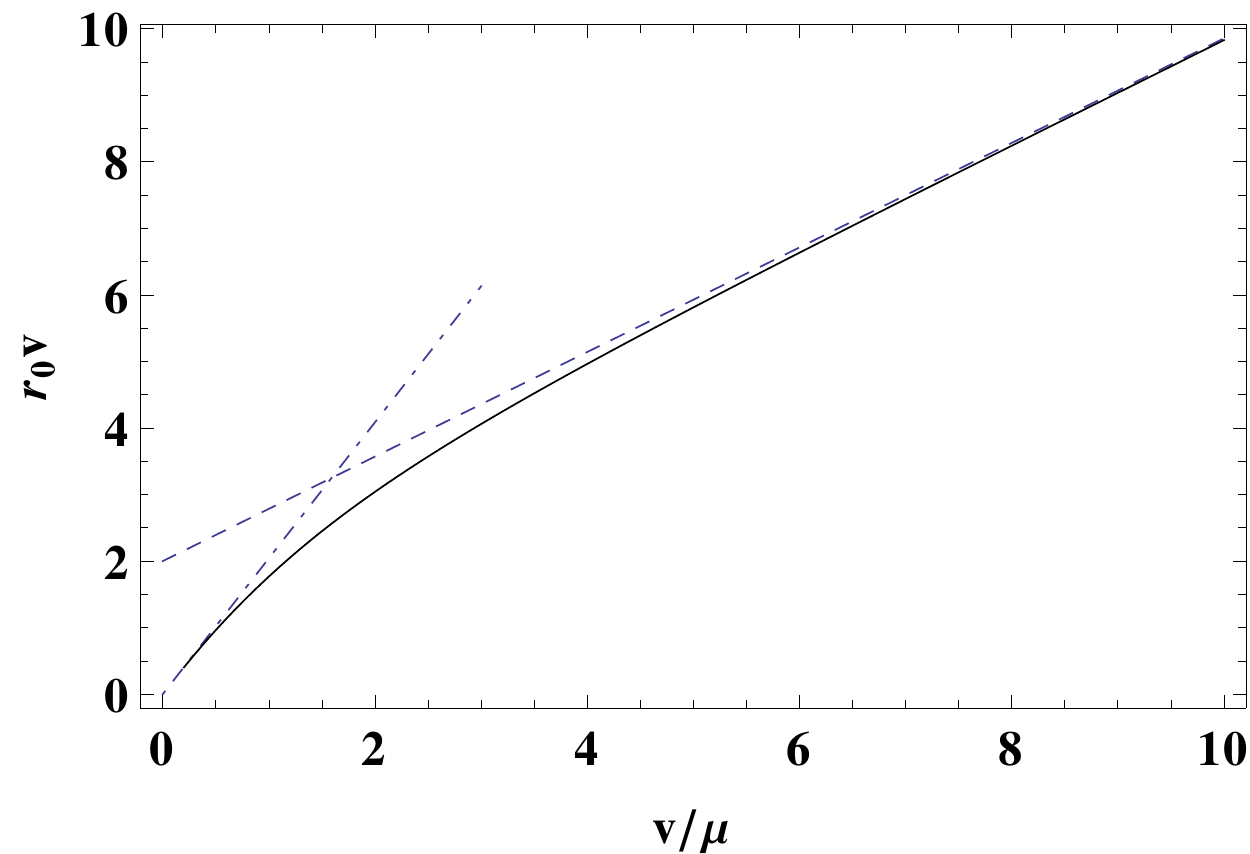} 
   \caption{The first zero $r_0$ of the density $\rho_{00}(r;\mu)$, as a function of $1/\mu$ (in units of $v$) -- black line. The dashed line and the dot-dashed line show the first zero for small $\mu$ given by \eqref{eq_small_mu} with $n=0$
   and for large $\mu$ given by $r_0=2.04279/\mu$, from \eqref{eq_large_mu}, respectively.}
   \label{fig:firstzero}
\end{figure}

For $\mu\ll v$ we can find these zeros explicitly by demanding that \eqref{eq:chipart} vanishes
\be
r_n\approx-\frac{1}{2\mu}\delta(\mu/v)+\frac{\pi (n+1)}{2\mu}\approx \frac{2}{v}+\frac{\pi }{4 \mu }+\frac{\pi n}{2\mu}\;,
\label{eq_small_mu}
\ee
where we expanded $\delta(\mu/v)$ for small $\mu$. For large $\mu$ we expect the first zero to be small, i.e. $rv\ll 1$, and get from \eqref{eq:dyonsol} that
\be
|\chi_1|^2-|\chi_2|^2\sim \frac{-2+(2-4\mu^2 r^2)\cos(2\mu r)+4\mu r \sin(2 \mu r)}{(r v)^2}
\label{eq_large_mu}
\ee
which has a zero for $r_0\mu=2.04279$. Fig.\ \ref{fig:firstzero} shows the numerical solution of the first zero, as well as the limits of $\mu/v\gg 1$ and $\mu/v\ll 1$.

Finally, we discuss the possibility of zero modes with physical, i.e.\ antiperiodic, boundary conditions. The static dyon discussed so far has normalizable zero modes for near periodic boundary phases, namely
\begin{align}
 \p\in\Big(-\frac{v}{2T},\frac{v}{2T}\Big)
\label{eq:range_static}
\end{align}
There is another embedding of the dyon into $S^1\times R^3$, that will play a role in the caloron solutions as well (as the complementary constituent dyon). To see how it arises note that the solution \eqref{eq:dyonsol} has the topological charge $Q=v/(2\pi T)$, which is fractional for $v/T\in[0,2\pi]$. In order to get a solution which complements the topological charge, we should replace the parameter $v\rightarrow \bar v=2\pi T-v$ and obtain a topological charge $\bar Q=1-Q$. However, this change of parameters alone is not sufficient to match the two configurations. For $3+1$-dimensional configurations it is useful to define the asymptotic Polyakov loop, also named holonomy,
\begin{align}
 \P\equiv\lim_{r\to\infty}\mathcal{P}\exp\left(i\int_0^{\b}\!\!\!dt\, A_0(t,\vec x)\right)
\label{eq:def_Ploop}
\end{align}
with $\mathcal{P}$ denoting path ordering along $x_0$. The starting configuration with gauge fields given by \eqref{eq:radialgaugefields_both} has 
\begin{align}
 \P=\exp\left(i\frac{v}{2T} \hat r\cdot\vec\tau\right)\,,\quad 
 \frac{1}{2}\,\tr\,\P=\cos(v/2T)
\label{eq:Ploop_dyon}
\end{align}
whereas the Polyakov loop of the new solution would be 
\be
\bP=\exp\left(i\frac{\bar v}{2T} \hat r\cdot\vec\tau\right)=-\exp\left(i\frac{v}{2T} \hat r\cdot\vec\tau\right)=-\P %-2\cos(v\beta/2)=-\P\;.
\ee
and thus different.

Therefore, consider the gauge transformation\footnote{In the gauge of the caloron, which combs the field $A_0$ to the third color direction, the associated gauge transform is $\Omega=\exp(i\pi \tau_3\,t/\b)$. For higher gauge group such gauge transformations need to obey $\Omega(t=\b)=z\,\Omega(t=0)$, where $z$ is an element of the center of the gauge group.}
\be\label{eq:gaugetrans}
 \Omega(t,\vec{x})=\exp\left(i\pi \hat{r}\cdot\vec{\tau}\,\frac{t}{\b}\right)
\ee
It is antiperiodic, $\Omega(\b,\vec{x})=-\one=-\Omega(0,\vec{x})$. As the gauge fields transform in the adjoint representation, they remain periodic. Hence, the dyon gauge field of (\ref{eq:radialgaugefields_both}) with the replacement $v\rightarrow \bar v$ transformed with $\Omega$ is a valid configuration on $S^1\times R^3$. The gauge transformation does not alter its action and topological charge, so they are still proportional to $\bar v$ (cf.~(\ref{eq:dyon_action_topcharge}), both are gauge invariant),  and thus also inherits (anti)selfduality; its gauge fields are just not static anymore. However, $\Omega$ does affect the Polyakov loop as $\bP\rightarrow -\bP=\P$. This means that the static dyon solution \eqref{eq:radialgaugefields_both} and the dyon solution with $v\rightarrow \bar v=2\pi T-v$ and the gauge twist of \eqref{eq:gaugetrans} both have the same asymptotic Polyakov loop $\P$ and complementary topological charges $Q, \bar Q=1-Q$ respectively.

In calorons of integer topological charges, these dyon solutions occur together (see below) and are sometimes refered to as ``light'' and ``heavy'' dyon. The reason for this nomenclature is because the Polyakov loop branch in full QCD, i.e. when fermions are dynamical,  interpolates between unity and (near) zero from high to low temperatures. This would correspond to the parameter $v$ in the range $v\in[0,\pi T]$. In this range the untwisted dyon (i.e. the one without the gauge transformation \eqref{eq:gaugetrans}) has the action and the topological charge proportional to $v$ and the one with the gauge twist has the same proportional to $\bar v$. Since in this range $\bar v\le v$ the dyons untwisted and twisted dyons are referred to as ``light'' and ``heavy''.

Finally, we note that the fundamental fermions feel the gauge transformation of \eqref{eq:gaugetrans}, they obtain an additional factor of $-1$ in their periodicity property. Consequently, the parameter $\p$ used in the ansatz \eqref{eq:ansatz} is not the boundary phase as defined in \eqref{eq:gaugetrans}. Rather, we should replace $\p\to \p-\pi=-\bp$ in this ansatz. With the additional replacement $v\to\bar v$ it follows that these zero modes on top of the twisted dyons are normalizable for
\begin{align}
 \bp\in\Big(-\frac{\bar{v}}{2T},\frac{\bar{v}}{2T}\Big)\rightleftharpoons \p\in\Big(\frac{v}{2T},2\pi-\frac{v}{2T}\Big)
\label{eq:range_nonstatic}
\end{align}
which is the complementary range of the static dyons, \eqref{eq:range_static}. For the chemical potential the replacement reads $\mu_\p\rightarrow \mu_{-\bar\p}=\mu-i\bar\p T$. Antiperiodic fermions are included as $\bar\p=0$ with $\mu$ remaining the same.

\subsection{Calorons and instantons}

In this section we will demonstrate that zero modes at imaginary chemical potential are included naturally in the ADHM-Nahm formalism \cite{Atiyah:1978ri,Christ:1978jy,Nahm:1979yw}, so that they are available -- at least in principle -- for all selfdual solutions of four-dimensional Yang-Mills theory.
After analytic continuation one thus obtains zero modes at real chemical potential. We will demonstrate this explicitly for SU(2) calorons and then in particular limits recover the zero modes of dyons from the previous section  
and of calorons and instantons from the literature.

\subsubsection{Calorons, especially in SU(2)}\label{sec:cal}

Calorons are (anti-)selfdual Yang-Mills fields (`instantons') at finite temperature $T$, i.e. over the space-time manifold $S^1\times R^3$ where the Euclidean time is compactified on a circle $S^1$ with circumference $1/T$.

The asymptotic Polyakov loop plays an important role as a parameter of caloron solutions. For the gauge group $SU(2)$ we parametrize it with the holonomy parameter $\omega\in[0,1/2]$:
\begin{align}
 \P=\exp(2\pi i\omega\tau_3)\,,\quad
 \frac{1}{2}\,\tr\,\P=\cos(2\pi\omega)\,.
\label{eq:def_Ploop_cal}
\end{align}
The two extreme cases $\omega=0,\,\omega=1/2$ amount to `trivial holonomy' $\P=\pm 1_2$, while the middle $\omega=1/4$ with $\tr\,\P=0$ is refered to as `maximally nontrivial holonomy'. From the corresponding asymptotics of the Polyakov loop, \eqref{eq:Ploop_dyon}, one can read off the connection to the (dimensionful) ``Higgs vev'' of \eqref{eq:higgs},
\begin{align}
 v=4\pi \omega T,\quad \omega=\frac{v}{4\pi T}\,.
\end{align}

As a new aspect in comparison to zero temperature instantons (which can be thought of as coming from dimensional reduction and symmetry breaking by $\P$), calorons in gauge group $SU(N_c)$ with topological charge $Q$ consist of $N_c|Q|$ constituent dyons. These appear as lumps of topological density with fractional topological charges given by the holonomy. In $SU(2)$ the topological charges are $2\omega$ (indeed, the identification above yields the monopole topological charge of \eqref{eq:dyon_action_topcharge}) and the complementary $1-2\omega\equiv2\bar{\omega}$. We will refer to the dyon with the higher topological charge as the ``heavy dyon'' and to the other as the ``light dyon''. The positions of them are moduli of the solution.  

The explicit gauge field of $SU(2)$ calorons have first been given in \cite{Kraan:1998pm,Lee:1998bb} (for multi-calorons see \cite{Bruckmann:2002vy}). We follow the ADHM inspired formalism of \cite{Kraan:1998pm}, reinstating the temperature $T$. The distance to the dyon of masses $\bar \omega$ and ${\omega}$ will be denoted by $r$ and $s$, respectively, and the distance between the dyons by $d=\pi\rho^2/\b$. In the limit of $\rho\to\infty$ a single dyon will be recovered. 

Note also, that the calorons in these references come in a particular gauge with two aspects. The first one, refered to as algebraic gauge, means that gauge fields are periodic up to the holonomy and $A_0\to 0$ far away from all the dyons. In other words, the holonomy $\P$ has been put into the twisted periodicity conditions. This is useful since such asymptotics simplify the superposition of dyons within the caloron and also the approximate superposition of dyons and antidyons in ensembles. The gauge transformation back to periodic gauge is given by $(\P)^{-t/\beta}$, which is non-periodic.

Secondly, the caloron is in a quasi-stringy gauge. For magnetic monopoles/dyons it is well-known that when one tries to ``comb'' the Higgs field $A_0$ to asymptotically approach a constant color direction, a singularity along a Dirac string appears. This is so because the winding of $A_0$, which equals the magnetic charge and is apparent in the hedgehog gauge, constitutes a topological obstruction against such a stringy gauge. In the caloron solutions, $A_0$ is proportional to $\tau_3$ almost everywhere; the corresponding winding caused by each dyon is performed in an exponentially small solid angle around the line connecting the dyons \cite{Bruckmann:2009pa}, cf.~\eqref{eq:auxfield} below. This quasi-stringy gauge  leads to large gradients along this line, but no singularities (apart from the caloron's center of mass, see \sec{sec:singledyonlim} and \sec{sec_numerics}).

\subsection{Profiles of caloron zero modes}
\label{sec_profiles_caloron}

The fermionic zero mode of the caloron  (at $\mu=0$) hops between the constituent dyons as a function of the periodicity angle. This has first been demonstrated in \cite{GarciaPerez:1999ux} and is in accordance with the discussion about dyon zero modes in \sec{sec:dyons}: for near periodic zero modes the static dyon supports the zero mode, whereas for the complementary range of boundary conditions around antiperiodic the complementary time-dependent dyon supports the zero mode. Index theorems on $S^1\times R^3$ can be found in \cite{Callias:1977kg,Nye:2000eg,Poppitz:2008hr}. This scenario will remain with real chemical potentials.
The explicit formulas for the zero modes are included in the ADHM formalism and are given by the ADHM Greens function $\hat{f}$ (see below). We will analytically continue them to obtain zero modes at real chemical potential.

In \cite{GarciaPerez:1999ux,Bruckmann:2003ag} explicit expressions for the zero mode with periodicity up to a phase and the holonomy (stemming from the algebraic gauge) 
\be
 \hat{\Psi}_{z/T}(t+\b,\vec{x})=e^{-2\pi i z/T}\P\hat{\Psi}_{z/T}(t,\vec{x})
\ee
were given (the dual variable to $z$ comes with $1/T$ since it has inverse length dimension). Gauging away $\P$ and comparing to (\ref{eq_period_second}) we identify $2\pi z/T=\p$, thus the zero mode with phase boundary conditions reads
\begin{align}
 \psii_0(x;\bar \p)&=(\P)^{-\t/\b}\hat{\Psi}_{\p/2\pi}(x)\nonumber\\ 
 &=(\P)^{-\t/\b}\hat{\Psi}_{1/2-\bar\p/2\pi}(x)
\end{align}
The zero mode at imaginary chemical potential follows by virtue of (\ref{eq_period_trafo})
\be
 \ps'_0(x;\bar\p)=(e^{i\bar\p}\P)^{-\t/\b}\hat{\Psi}_{1/2-\bar\p/2\pi}(x)\,.
\ee
Finally, the zero mode at real chemical potential follows through analytic continuation, i.e.\ the replacement $\bar\p=i\mu/T$,
\be
 \ps_0(x;\mu)=e^{\mu \t}(\P)^{-\t/\b}\hat{\Psi}_{1/2-i\mu/2\pi T}(x)
\ee
which is the antiperiodic mode.
Altogether, the results from \cite{GarciaPerez:1999ux,Bruckmann:2003ag} have to be taken at a complex $z=T/2-i\mu/2\pi$ and multiplied by a real factor and color phases from $\P=\exp(2\pi i\omega\tau_3)$. 
We obtain 
\begin{widetext}
\begin{align}\label{eq:fermsol}
 \!\!\ps_0(\mu)^A_\alpha=\rho\frac{\sqrt{\phi}}{2\pi}e^{\mu t}\Big[e^{-2\pi i  \omega\, tT\tau^3}\Big\{
  \partial_\nu\hat f\left(\omega T,\frac{T}{2}-i\frac{\mu}{2\pi}\right)
    \left(\begin{array}{cc}1&\\&0\end{array}\right)
  +\partial_\nu\hat f\left((1-\omega) T,\frac{T}{2}-i\frac{\mu}{2\pi}\right)
    \left(\begin{array}{cc}0&\\&1\end{array}\right)\Big\}
  \bar{\sigma}_\nu\epsilon\Big]_{A\alpha}\;,
\end{align}
with the following auxiliary functions
\begin{subequations}\label{eq:componentform}
\begin{align}
 \phi=&\,\frac{ \psi}{\hat\psi}\;,\qquad
 \hat\psi=-c_t+c_r c_s+\frac{r^2+s^2-d^2}{2rs}\,s_r s_s\\
\psi=&-c_t+c_r c_s+\frac{r^2+s^2+d^2}{2rs}s_r s_s+d\Big(s^{-1}s_s c_r+r^{-1}s_r c_s\Big)\\
c_t=&\cos(2\pi t T)\,,\quad c_r=\cosh(4\pi r\bar\omega T)\,,\quad c_s=\cosh(4\pi s\omega T)\,,\quad s_r=\sinh(4\pi r\bar\omega T)\,,\quad s_s=\sinh(4\pi s\omega T)\\
\label{eq:ffunct}
\hat f(z,z')=&\,e^{2\pi i t(z-z')}\pi T(rs\psi)^{-1}\Big\{e^{-2\pi i t T\; \text{sign}(z-z')}\sinh(2\pi r|z-z'|)s\nonumber\\
&-r^{-1}\cosh(2\pi r(z+z'-T))\Big[d s c_s+\frac{1}{2}(s^2-r^2+d^2)s_s\Big]\nonumber\\
&+r^{-1}\cosh(2\pi r(2\bar\omega T-|z-z'|))\Big[d s c_s+\frac{1}{2}(r^2+s^2+d^2) s_s\Big]\nonumber\\
&+\sinh(2\pi r(2\bar\omega T-|z-z'|))\Big[s c_s+d s_s\Big]\Big\}
\end{align}\end{subequations}
We repeat that $r$ and $s$ are the distances of $\vec{x}$ to the dyon centers and $d$ is the distance between the dyons. The caloron gauge field is given in terms of these functions, too \cite{Kraan:1998pm}. 
The Greens function at the arguments needed simplifies to  \enlargethispage{-3\baselineskip}
\begin{align}
 \hat f(\omega T,\frac{T}{2}-i\frac{\mu}{2\pi})=&\,e^{-\pi i tT}e^{2\pi i\omega tT}e^{-\mu t}\pi T(rs\psi)^{-1}\Big\{e^{2\pi i t T\; }\sinh(\pi r T -2\pi r \omega T-i\mu r)s\nonumber\\
&+\cosh(\pi r T -2\pi r \omega T+i\mu r)r s_s+\sinh(\pi r T -2\pi r \omega T+i\mu r)\Big[s c_s+d s_s\Big]\Big\}\\
 \hat f((1-\omega) T,\frac{T}{2}-i\frac{\mu}{2\pi})
=&\,\hat f(\omega T,\frac{T}{2}-i\frac{\mu}{2\pi})^*
\label{eq_cc}
\end{align}
The last relation reflects the anti-unitary symmetry \eq{anti_symm_cont} of SU(2) and has also been used (at $\mu=0$) in \cite{GarciaPerez:1999ux} to relate the components of the zero mode.

\begin{figure*}[!t]
\begin{center}
\includegraphics[width=0.32\textwidth]{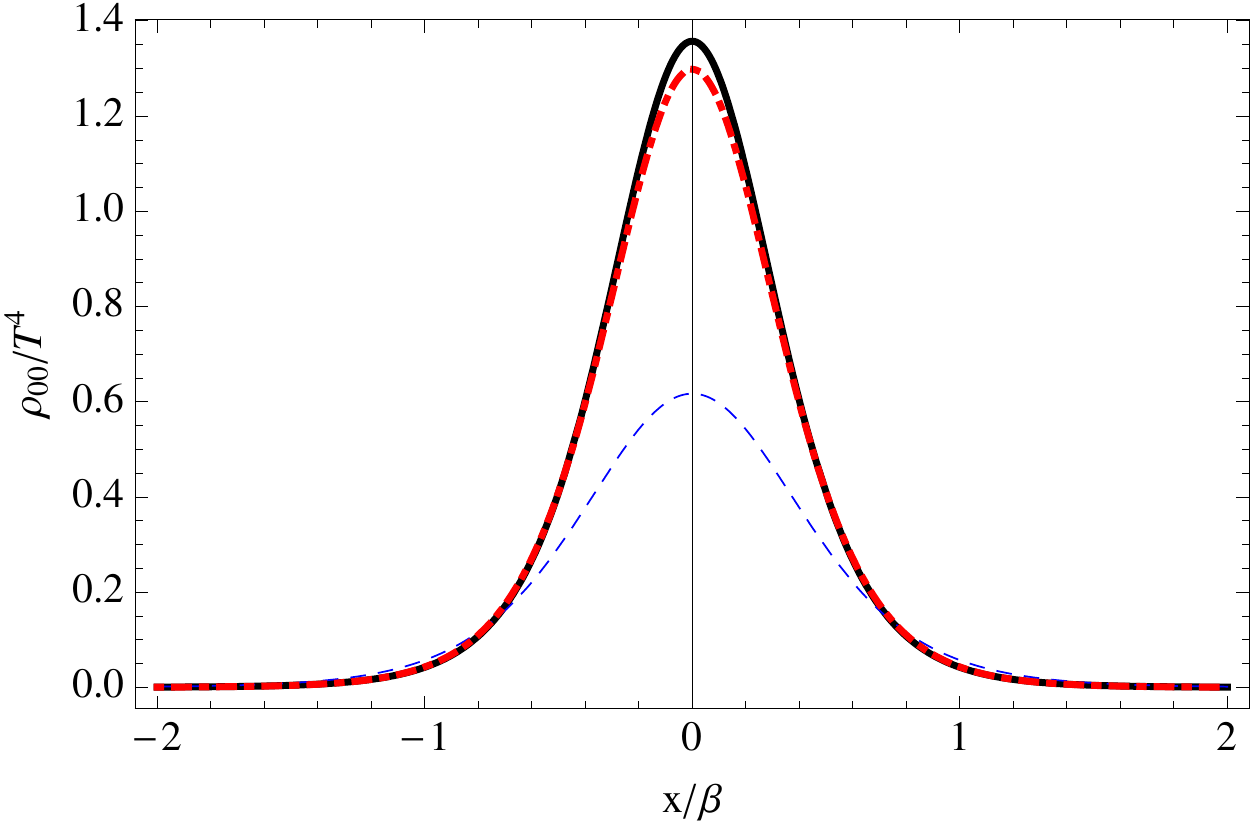}\:
\includegraphics[width=0.32\textwidth]{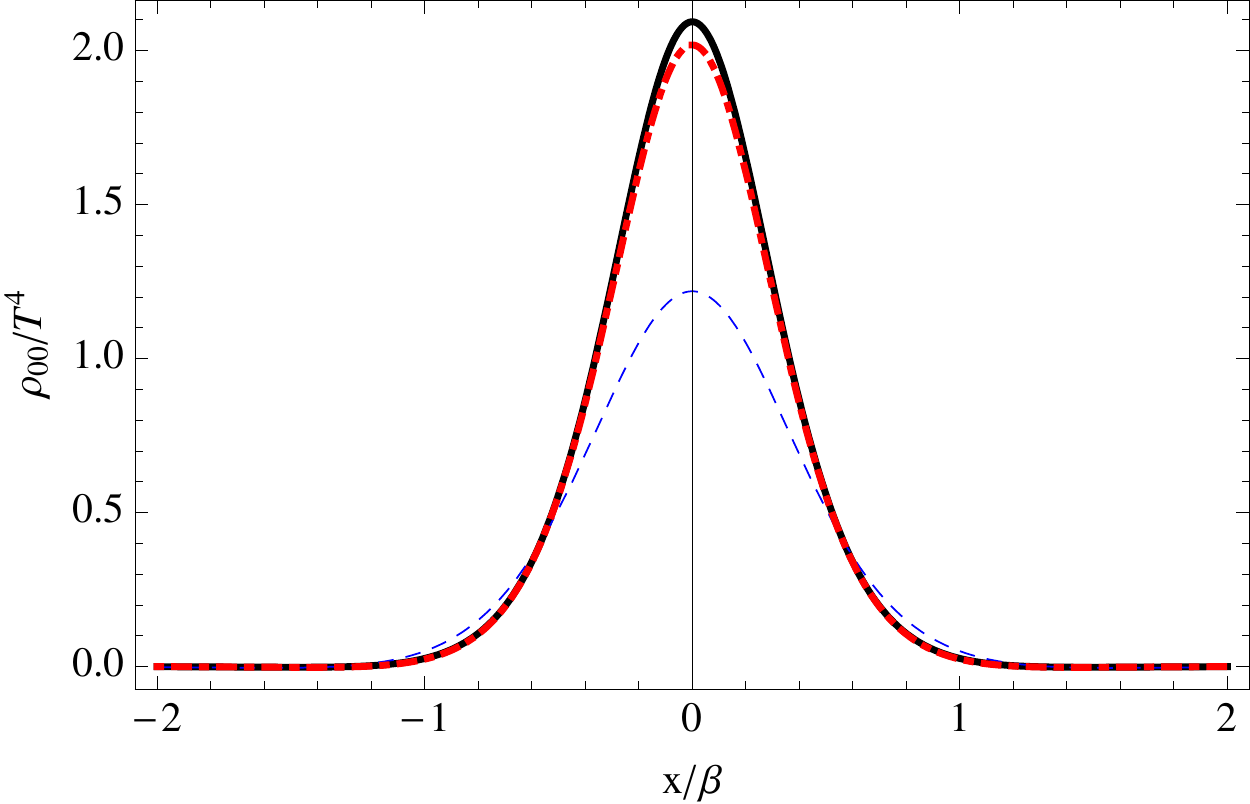}\:
\includegraphics[width=0.32\textwidth]{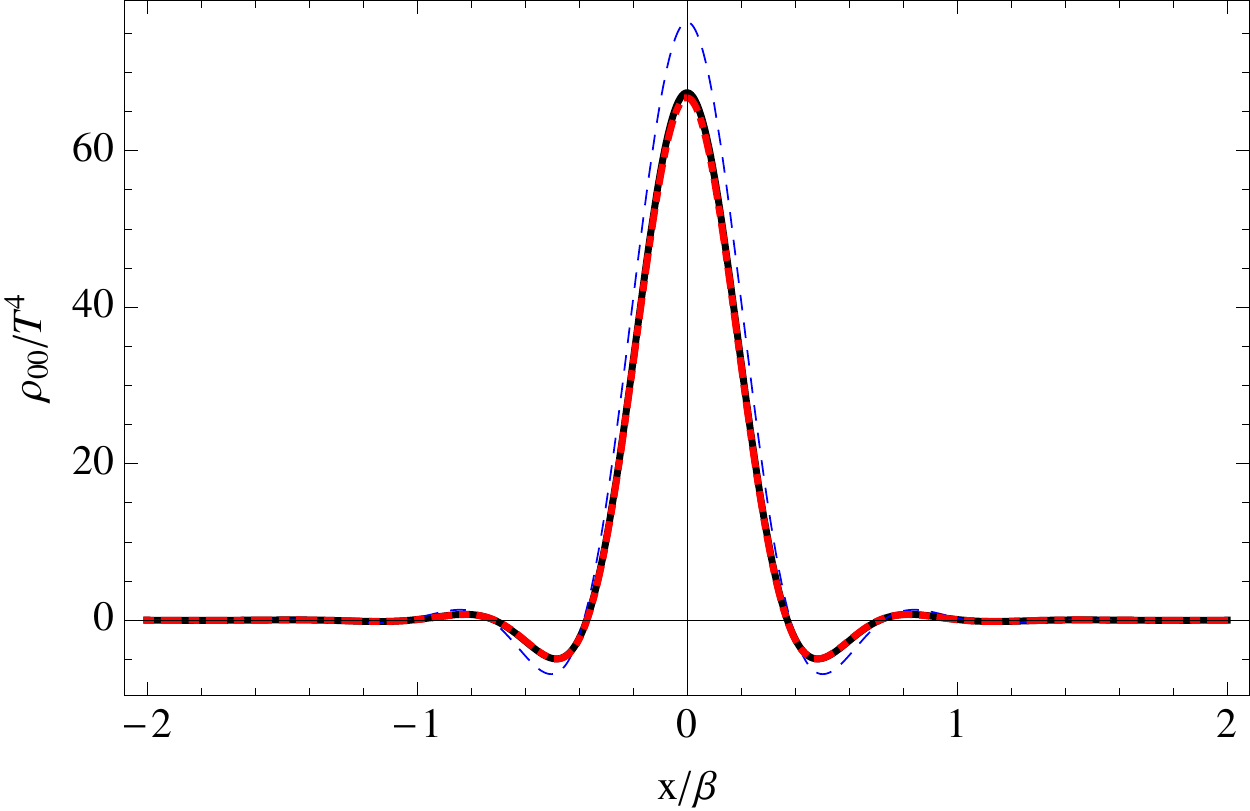}
\caption{Densities of zero modes of a caloron (solid black line), single dyon (dashed blue), and single dyon with correction to the holonomy $v\rightarrow v+1/d$, where $d$ is the distance to the other dyon (dot-dashed red). The plots are given for 3 values of $\mu/T=0,1,5$ from left to right. The densities are along the axis transverse to the position of the second dyon in the caloron. All plots are given for $\rho=0.6\beta$ and for maximally nontrivial holonomy $\omega=\bar\omega=1/4=v/(4\pi T)= \bar v/(4\pi T)$.}\label{fig:calzero mode}
\end{center}
\end{figure*}

For the density a simpler formula applies
\begin{align}
 \rho_{00}(\mu)=-\frac{1}{(2\pi)^2}\,\partial_\nu^2 \hat f
\label{eq:density00}
\end{align}
with the Greens function
\begin{align}\label{eq:ffunctwithmu}
 \hat f\equiv 
 \hat f\left(\frac{T}{2}-i\frac{\mu}{2\pi},\frac{T}{2}-i\frac{\mu}{2\pi}\right)=\,\pi T (rs\psi)^{-1}\Big\{&-r^{-1}\cos(2\mu r)\Big[d s c_s+\frac{1}{2}(s^2-r^2+d^2)s_s\Big]\nonumber\\
 &+r^{-1} c_r\Big[d s c_s+\frac{1}{2}(r^2+s^2+d^2) s_s\Big]+s_r \Big[s c_s+d s_s\Big] \Big\}
\end{align}
\end{widetext}
The reader will note that the function $\hat f(z,z')$ given above is not the full function for arbitrary arguments $z,z'$. The full function can be found in the original papers by Kraan and van Baal \cite{Kraan:1998pm}, we have given only the relevant part for our discussion valid for $z/T,z'/T\in[\omega,1-\omega]$.

The resulting zero mode densities are shown in Fig. \ref{fig:calzero mode}, where we also compare with the single dyon zero mode discussed previously. Again these densities have negative regions. Note an interesting effect that even for fairly large values of the size parameter $\rho$, i.e.\ for well-separated dyons, the two densities do not match very well. This mismatch can be compensated for by a trivial adjustment to the effective holonomy of a single dyon. Namely the solution for a single dyon with holonomy $v$ should be changed to that with holonomy $v+1/d$ in order to match the solution in the caloron. In other words the holonomy at the dyon centre is not $v$, but is corrected by the asymptotic value of the other constituent of caloron and is approximately $v+1/|\vec d+\vec r|\approx v+1/d$. This has been noticed in the full solution \cite{Bruckmann:2003ag} where it was shown that single dyon ``mass'' is renormalized by the presence of the other dyon. This renormalization of mass is not new and has been know for a long time for Yang-Mills+adjoint Higgs system and that it effects the classical interactions of monopoles \cite{Manton:1977er,Goldberg:1978ee,Magruder:1978br,O'Raifeartaigh:1979ca}. Fig. \ref{fig:calzero mode} also shows the solutions with this adjustment. Considering the crudeness of this approximation, the agreement is remarkable up to fairly small instanton size.

\subsection{Various limits}

In this section we will discuss various limits of the general caloron solution of \eqref{eq:fermsol} and (\ref{eq:density00},\ref{eq:ffunctwithmu}). In particular, we take the trivial holonomy limit and compare it to previous work, as well as the  zero temperature limit and the large separation limit, which leads to single dyons. These limits are concisely summarized in Table \ref{tab:limits}.

\subsubsection{Trivial holonomy limit $\omega\rightarrow 0$}

In the trivial holonomy limit $\omega\to 0,\,\bar\omega\to 1/2$ the auxiliary functions simplify in the following way
\begin{align}
 c_r&=\cosh(2\pi rT)\,,\quad c_s=1\,,\nonumber\\
 \quad s_r&=\sinh(2\pi rT)\,,\quad s_s=0\,,
\end{align}
immediately leading to (writing out $d=\pi\rho^2 T$)
\begin{align}
 \hat\psi&=-c_t+c_r\,,\quad\psi=-c_t+c_r+\frac{\pi\rho^2 T}{r}\, s_r\,,\nonumber\\
 \phi&=1+\frac{\pi\rho^2T}{r}\frac{s_r}{c_r-c_t}=\Pi
\end{align}
where we have identified $\phi$ with the function $\Pi$ used in the literature \cite{Grossman:1977mm,Schafer:1996wv,Cristoforetti:2011fq}.

Concerning the Greens functions \eqref{eq:ffunct}, 
all dependence on $s$ drops as it should to restore spherical symmetry. From \eqref{eq:ffunctwithmu} $\hat f$ we obtain
\begin{align}
 \hat f &=\frac{\pi T}{r\psi}\Big\{\frac{d}{r}\big(-\cos(2\mu r) + c_r\big)+s_r\Big\}\\
 &=\frac{\pi^2\rho^2T^2}{r^2}\frac{-\cos(2\mu r)+\cosh(2\pi rT)+\frac{r}{d}\sinh(2\pi rT)}{-\cos(2\pi t T)+\cosh(2\pi rT)+\frac{d}{r}\sinh(2\pi rT)}\nonumber
\end{align}
We recall that the zero mode density follows from acting with the four-dimensional Laplacian on this function, \eqref{eq:density00}.

For the zero mode components, the corresponding Greens function simplifies to 
\begin{align}
 \hat f(0, \frac{T}{2}-i\frac{\mu}{2\pi})
  =&\,\frac{\pi T}{r\psi}e^{-\mu t}\big\{e^{i\pi t T}\sinh(\pi r T-i\mu r)\nonumber\\
 &+e^{-i\pi t T}\sinh(\pi r T+i\mu r)\big\}\nonumber\\
 =&\,\frac{2\pi T}{r\psi}e^{-\mu t}\big\{\!\cos(\mu r)\cos(\pi tT)\sinh(\pi r T)\nonumber\\
 &+\sin(\mu r)\sin(\pi tT)\cosh(\pi r T)\big\}
\end{align}
This agrees with $\hat f(T,T/2-i\mu/2\pi)$ because the Greens function $\hat f(z,z')$ is periodic in both arguments by construction \cite{Kraan:1998pm}, and also because both are real using \eq{eq_cc}.
The projection matrices in \eqref{eq:fermsol} then add up to the unity matrix. Put together, the zero mode reads
\begin{align}
 \ps_0(\mu)^A_\alpha=\frac{\rho}{2\pi}e^{\mu t}\sqrt{\Pi}\,\partial_\nu\hat f\left(0,\frac{T}{2}-i\frac{\mu}{2\pi}\right)
  (\bar{\sigma}_\nu\epsilon)_{A\alpha}
\end{align}
To compare to previous work we write
\begin{align}
 &\hat f(0, \frac{T}{2}-i\frac{\mu}{2\pi})
  =\frac{e^{-\mu t}}{\rho^2}\frac{\Phi}{\Pi}\,,\\
 &\Phi=(\Pi-1)\Big\{\frac{\cos(\mu r)\cos(\pi tT)}{\cosh(\pi r T)}+\frac{\sin(\mu r)\sin(\pi tT)}{\sinh(\pi r T)}\Big\}\nonumber
\end{align}
which turns the zero mode into the form given in  \cite{Cristoforetti:2011fq}, for this case of trivial holonomy.

\subsubsection{Instanton limit $T\rightarrow 0$}

In the zero temperature limit, calorons become instantons irrespective of the original holonomy. Indeed, all $\omega$ and $\bar \omega$ enter with $T$ such that they disappear in this limit. From the previous trivial holonomy case we get
\begin{equation}
\hat f=\frac{1}{r^2+t^2+\rho^2}\left(1+\frac{\sin^2(\mu r)\rho^2}{r^2}\right)
\end{equation}
and 
\begin{align}
 \Pi&=1+\frac{\rho^2}{r^2+t^2}\,,\\
 \Phi&=\frac{\rho^2}{r^2+t^2}\,\big(\cos(\mu r)+\frac{t}{r}\sin(\mu r)\big)\,.
\end{align}
which agrees with \cite{AragaodeCarvalho:1980de,Abrikosov:1983rh,Cristoforetti:2011fq}.
Note that, as expected, the time and space coordinate go together forming the four-dimensional radius almost everywhere apart from the $\mu$-dependent part. The corresponding zero mode densities also contain negative region with zeroes separated again by $\Delta r =\pi/(2\mu)$ and an increasing value at the core, $\ch_{00}(t=0,r=0;\mu)\sim(1/\rho^2+\mu^2)^2$.  

\begin{table*}[!hbt]
 \begin{tabular}{|c||c|c|c|}\hline
  &$\omega\rightarrow 0$ & $T\rightarrow$0 &$\rho\rightarrow\infty$\\
 \hline\hline
$\hat f$&$\frac{\pi^2\rho^2 T^2}{r^2\psi}\left(-\cos(2r\mu)+\cosh(2\pi rT)\right)+\frac{\pi T}{r\psi}$ & $\frac{1}{r^2+t^2+\rho^2}\left(1+\frac{\sin^2(r\mu)\rho^2}{r^2}\right)$ &$\frac{\pi T}{r}\frac{\cosh(4\pi rT\bar\omega)-\cos(2r\mu)}{\sinh(4\pi r T\bar\omega)}$\\\hline
$\psi$ & $-\cos(2\pi tT)+\cosh(2\pi rT)+\frac{\pi\rho^2 T}{r}\sinh(2\pi rT)$ &$2\pi^2 T^2(r^2+\rho^2)$&$\frac{\pi\rho^2T}{r}\sinh(4\pi r T\bar\omega)e^{4\pi^2\rho^2T^2\omega}$\\\hline
$\hat\psi$&$-\cos(2\pi tT)+\cosh(2\pi rT)$ &$2\pi^2T^2r^2$ &$\Big[\cos\theta\sinh(4\pi r\bar\omega)+\cosh(4\pi r\bar\omega)\Big]e^{4\pi^2\rho^2T^2\omega}$\\\hline
$\phi$&$1+\frac{\pi\rho^2 T}{r}\frac{\sinh(2\pi r T)}{-\cos(2\pi t T)+\cosh(2\pi r T)}$&$1+\frac{\rho^2}{t^2+r^2}$&$\rho^2{\frac{2\pi T}{r}}\frac{1}{{\cos\theta+\coth(4\pi r\bar\omega T)}}$\\\hline 
$\hat f_{1-\omega}$&$\frac{2\pi T}{r\psi}e^{-\mu t}\Big[\cos(\pi t T)\sinh(\pi r T)\cos(r\mu)$&&\\
&\hspace{1.2cm}$+\sin(\pi t T)\sin(r\mu)\cosh(\pi rT)\Big]$&$\frac{2\pi^2 T^2}{\psi}e^{-\mu t}\Big[ \cos(r\mu)+\frac{t}{r}\sin(r\mu) \Big]$&$\frac{1}{2\rho^2}e^{-\mu t}e^{2\pi i t\bar\omega}\Big[\frac{\cos(r\mu)}{\cosh(2\pi r T\bar\omega)}-\frac{i\sin(r\mu)}{\sinh(2\pi r T\bar\omega)}$\\\hline
$\hat f_\omega$&$\hat f_{1-\omega}$&$\hat f_{1-\omega}$&$\frac{1}{2\rho^2}e^{-\mu t}e^{-2\pi i t\bar\omega}\Big[\frac{\cos(r\mu)}{\cosh(2\pi r T\bar\omega)}+\frac{i\sin(r\mu)}{\sinh(2\pi r T\bar\omega)}\Big]$\\\hline
 \end{tabular}
\caption{Table summarising some limits of functions needed to determine the zero mode density $\psi_0^\dagger(-\mu)\psi_0(\mu)=-\frac{1}{4\pi^2}\partial_\mu\partial^\mu\hat f$, and the component form given by \eqref{eq:componentform}. We labeled $\hat f_{\omega}=\hat f_{\omega}(\omega T,T/2-i\mu/2\pi)$.}\label{tab:limits}
\end{table*}

\subsubsection{Single dyon limit $\rho\rightarrow\infty$}\label{sec:singledyonlim}

The limit of a single dyons can be obtained in the limit of large size $\rho$, i.e.\ large separation $d=\pi\rho^2T$, of the dyon constituents. While $r$ remains finite, $s$ is large, too, $s=\sqrt{d^2+r^2-2dr\cos\theta}=d-r\cos\theta+\mathcal{O}(d^{-1})$, where $\theta$ is the angle between $r$ and the line connecting the dyons, its appearance originates from the quasi-stringy gauge of the caloron.

The Greens function needed for the profile becomes static,
\be
\hat f=\frac{\pi T}{r}\frac{\cosh(4\pi rT\bar\omega)-\cos(2\mu r)}{\sinh(4\pi rT\bar\omega)}
\ee
Here, the auxiliary functions $c_s$ and $s_s$ have become exponentially large in $d$ and suppress the time-dependence in $\psi$. Furthermore, $c_s=s_s$ up to exponentially small corrections and these factors have canceled in all expressions. The three-dimensional Laplacian yields the profile from it via \eq{eq:density00}. It can be shown that this precisely equals the density \eqref{eq:alphas} of the dyon zero mode upon identifying $4\pi T \bar\omega=\bar v$. We remind the reader that we started with the antiperiodic zero mode for the caloron and thus arrive at the antiperiodic zero mode for a time-dependent dyon. Therefore, its vev needs to be $\bar v$; the  transform $\Omega$ discussed in \sec{sec_analytic} is a gauge transform and therefore not visible in the fermionic profile, only in the components to be discussed now.

The caloron's auxiliary function 
\begin{align}\label{eq:auxfield}
 \phi=d\cdot \frac{2}{r}\,\frac{1}{\coth(4\pi r\bar\omega T)-\cos\theta}
\end{align}
is linearly divergent in $d$ and shows the quasi-stringy behaviour \cite{Bruckmann:2003ag}: for large $r$ the denominator would be zero on the half-line $\theta=0$, if there were no exponentially small deviations of the $\coth$-term from 1. The factor $\rho\sqrt{\phi}$ in the zero mode is $\mathcal{O}(d)=\mathcal{O}(\rho^2)$.
The corresponding limit of the Greens functions are
\begin{align}
 \lim_{\rho\to\infty}
 \rho^2 \hat f(\omega T,&\frac{T}{2}-i\frac{\mu}{2\pi})
 =
 \,e^{-2\pi i \bar{\omega} tT}e^{-\mu t}\\
 &\times\frac{1}{s_r}\sinh(2\pi \bar{\omega} r T+i\mu r)\nonumber\\
 \lim_{\rho\to\infty}
 \!\!\rho^2 \hat f((1-\omega) T,&\frac{T}{2}-i\frac{\mu}{2\pi})
 =
 \,e^{2\pi i \bar{\omega} tT}e^{-\mu t}\\
 &\times\frac{1}{s_r}\sinh(2\pi \bar{\omega} r T-i\mu r)\nonumber\\
 \lim_{\rho\to\infty}
 \rho^2 \hat f( \left(\substack{
\omega\\1-\omega}\right) T,&\frac{T}{2}-i\frac{\mu}{2\pi})
 =\frac{1}{2}\,e^{\mp 2\pi i \bar{\omega} tT}e^{-\mu t}\\
 &\times\Big(\frac{\cos(\mu r)}{\cosh(2\pi r\bar\omega T)}\pm i\frac{\sin(\mu r)}{\sinh(2\pi r\bar\omega T)}\big)\nonumber
\end{align}
The projection matrices can be written in terms of the identity and the third Pauli matrix, such that the zero mode reads
\begin{align}
 \psi_0(\mu)_\alpha^A=&\,\frac{\sqrt{T}}{2\sqrt{2\pi r}}\frac{1}{\sqrt{\coth(4\pi r\bar\omega T)-\cos\theta}}\,e^{\mu t}\Big[e^{-2\pi i \omega t T\tau_3}\nonumber\\
 &\times\partial_\nu \Big\{e^{-\mu t}\Big(\frac{\cos(\mu r)}{\cosh(2\pi r\bar\omega T)}+i\tau_3\frac{\sin(\mu r)}{\sinh(2\pi r\bar\omega T)}\Big)\nonumber\\
 &\qquad\:\:\times e^{-2\pi i \bar \omega t T\tau_3}\Big\} \bar \sigma_\nu\epsilon\Big]_{A\alpha}
\label{eq:psi0_single}
\end{align}
Performing the time and spatial derivatives this gives
\begin{align}
 &\psi_0(\mu)_\alpha^A=\frac{\sqrt{T}\pi\bar\omega T}{2\sqrt{2\pi r}} e^{-\pi i tT\tau_3}Q\\
 &\times\Big[\Big(\frac{\sin(\mu r)}{\sinh(2\pi r\omega T)}\coth(2\pi r\omega T)-\frac{\mu}{2\pi \bar\omega T}
 \frac{\cos(\mu r)}{\cosh(2\pi r\omega T)}\Big)\nonumber\\
 &\quad\:-i\Big(\frac{\cos(\mu r)}{\cosh(2\pi r\omega T)}+\frac{\mu}{2\pi \bar\omega T}\frac{\sin(\mu r)}{\sinh(2\pi r\omega T)}\Big)\hat r\cdot\vec{\tau}\epsilon\Big]_{A\alpha}\nonumber\\
 &Q=\frac{\tau_3\hat r\cdot\vec{\tau}+\tanh(2\pi r\bar\omega T)}{\sqrt{\coth(4\pi r\bar\omega T)-\cos\theta}}
\end{align}
This is the full antiperiodic zero mode of a dyon in the quasi-stringy gauge. The explicit appearance of $\tau_3$ and the angle $\theta$ renders it different from the hedgehog gauge representation used in \sec{sec:dyons}. These only enter $Q$, and $Q$ can be shown to be proportional to a (space-dependent) $SU(2)$ matrix $\mathcal{U}$
\begin{align}
 Q=2\,\frac{\sinh(2\pi r\bar\omega T)}{\sqrt{\sinh(4\pi r\bar\omega T)}}\,\mathcal{U}(r,\theta)
\end{align}
Performing a local (and periodic) gauge transformation the zero mode in hedgehog gauge finally becomes
\begin{align}
 &(e^{-\pi i tT\tau_3}\,(i\hat r\cdot\vec{\tau})\,\mathcal{U}^\dagger\,e^{\pi i tT\tau_3})\psi_0(\mu)=\pi\bar \omega T^{3/2}\frac{e^{-\pi i tT\tau_3}}{\sqrt{2\pi r s_r}}\nonumber\\
 &\times \Big[\Big(\cos(\mu r)\tanh(2\pi r \bar \omega T)+\frac{\mu}{2\pi \bar\omega T}\sin(\mu r)\Big)\\
 &\quad\:-i\Big(-\sin(\mu r)\coth(2\pi \bar\omega T)+\frac{\mu}{2\pi \bar\omega T}\cos(\mu r)\Big)\hat r\cdot\vec{\tau}\Big]\epsilon \nonumber
\end{align}
This solution matches the one of the time-dependent dyon discussed at the end of \sec{sec:dyons}.

\section{Numerical results for the staggered Dirac operator in caloron backgrounds}
\label{sec_numerics}

In this section we will compare the caloron zero modes from the continuum to lattice zero modes of the staggered Dirac operator.
The dimension of our lattices will be denoted by $N_x\cdot N_y \cdot N_z \cdot N_t$, the sites by $an$ with $a$ being the lattice spacing and 
normalized lattice vectors in the $\nu$th direction by $\hat{\nu}$.
All observables of dimension energy will be measured in units of temperature $T= 1/\beta=1/(N_t a)$, all length scales are given in units of $\beta$.

\subsection{Discretization of the caloron incl. smearing}
\label{subsec_discretization}

The first task is to compute the lattice links $U_\nu(an)$ from the continuum gauge fields of calorons. To this end we calculate 
gauge transporters
by discretizing the path ordered exponentials,
\begin{align}
&\mathcal{P} \exp\Big[i\int\limits_{an}^{a(n+\hat{\nu})} A_{\nu}(x) dx_{\nu} \Big] \nonumber \\
&\cong  \prod\limits_{k=1}^N \exp\left\{i \frac{a}{N}A_\nu\left[a\left(n + \frac{k}{N}\hat{\nu}\right)\right] \right\} \equiv U_\nu(an).
\label{eq:gaugelink}
\end{align}
The gauge field $A_\nu(x)$ is singular at the 
caloron's center of mass and has large gradients 
on the line connecting the constituent dyons 
as one can see from Fig. 1 in \cite{Bruckmann:2002vy} and Fig. 2 in \cite{Gerhold:2006bh}.
Therefore, the number $N$ of discretization points is adapted locally in steps of 40 until
both the real and imaginary part of all matrix elements of the links $U_\nu(an)$ change by less than $10^{-4}$ (between $N$ and $3N/2$). 
The typical value of $N$ is 40 except at the line which connects the dyons.

With this technique we obtain for a caloron with holonomy parameter $\omega=1/4$ and size parameter $\rho= \frac{3}{4}\beta$ on a $24\cdot24 \cdot 36 \cdot 8$ 
lattice $184\%$ of the continuum action $S_{cont}$. 
The additional action originates from the ``spatial boundary of the lattice'', where onto the finite box we force the infinite volume calorons, whose gauge fields are at variance with the periodic spatial boundary conditions.
To remove these artefacts we apply APE smearing \cite{Falcioni1985,Forcrand2007} with 
a parameter $\alpha_{APE}=0.45$. In Ref.~\cite{DeGrand:1997ss} it was shown that this value provides 
the best matching between APE-smearing and RG-cycling. 
Moreover, in Ref.~\cite{Ilgenfritz:2006ju} APE smearing with this value has been applied to SU(2) Monte-Carlo gauge configurations to reveal their dyon content. 

After $275$ APE smearing steps the boundary artefacts are sufficiently smoothed out reaching now $1.07\%$ of $S_{cont}$.
In order to have comparable results all gauge links are smeared until the lattice action reaches $1.07\, S_{cont}$.

\subsection{Staggered Dirac operator at finite chemical potential and its symmetries}
\label{subsec_staggered_def}

We use the standard\footnote{Improvements are not necessary on our smooth backgrounds.} staggered operator, which in our notation has the form
\begin{align}
 D(\mu;n|m) = \frac{1}{2a} \sum\limits_{\nu=1}^4 \eta_{\nu}(n) \Big\{&
  \delta_{n + a\hat{\nu}, m}  U_{\nu}(an) e^{a \mu \delta_{\nu,4}}\\ \nonumber 
  -& \delta_{n - a\hat{\nu}, m} U^{\dagger}_{\nu}(am) e^{- a\mu \delta_{\nu,4}}\Big\}. 
\end{align}
where $\eta_{\nu}(n)=(-1)^{n_1+\ldots n_{\nu-1}}$ is the staggered sign. 
At vanishing chemical potential $D$ is anti-hermitian with purely imaginary eigenvalues as in the continuum, whereas at nonzero chemical potential the hermiticity relation \eqref{eq_related_operator} to opposite $\mu$ holds. We have implemented the chemical potential by exponential factors on all temporal links.

This operator obeys a remnant of chiral symmetry
\begin{align}
\{D(\mu),\eta_5\} = 0 \qquad \eta_5(n)=(-1)^{n_1+\ldots n_4}
\label{eq:staggered_chiral_symmetry}
\end{align} 
yielding eigenvalues $\pm\l$ as in the continuum. The anti-unitary symmetry, however, differs from the one of the continuum Dirac operator\footnote{because no charge conjugation matrix is needed for the staggered operator being a scalar in spin spcae}. It reads 
\begin{align}
 %U_{\nu}^*=\tau_2^\dagger U_{\nu} \tau_2 \quad \Rightarrow 
 \quad D^{*}(\mu) = \tau_2^\dagger D(\mu) \tau_2
\label{eq:pseudoreality}
\end{align}
with $\tau_2^*\tau_2=-1$. The staggered Dirac operator is thus in the universality class of the Gaussian symplectic 
ensemble (the expected transition to the continuum symmetry has been investigated in \cite{Follana:2006zz,Bruckmann:2008xr}). 
As a consequence, $\tau_2\ps^*$ is another eigenmode with eigenvalue $\lambda^*$. Together with the chiral 
symmetry, \eqref{eq:staggered_chiral_symmetry}, this yields (Kramer's) degenerate eigenvalues for vanishing chemical potential. 
At nonzero chemical potential the eigenvalues come in quartets $\{\lambda,-\lambda^*,-\lambda,\lambda^*\}$ with eigenmodes $\{\ps,\eta_5\tau_2\ps^*,\eta_5\ps,\tau_2\ps^*\}$, respectively. 

On our smooth backgrounds an approximate four-fold degeneracy is expected from the four tastes of the staggered operator. Thus, instead of each exact continuum zero mode a low-lying quasi-zero quartet shall occur. At $\mu=0$ topological quasi-zero modes of the staggered operator are known to exist. The chirality is represented by the four-link operator $\Gamma_{50}$, with which these modes have expectation values close to $\pm 1$, in contrast to higher modes \cite{Smit:1986fn,Pugh:1989qd,Laursen:1990ec,Durr:2013gp}.
Both, nonzero real parts and nonzero imaginary parts of the eigenvalues could be induced by discretization and finite volume effects. They may also differ for quasi-zero eigenvalues and singular values.

On the smeared configurations we find modes \textit{separated by orders of magnitude} from the rest of the spectrum (see Fig. \ref{fig:spectrum}). 
We investigate the profiles of these quasi-zero modes without further analysis of their continuum and infinite volume limit.
As a side remark we state that on a $48\cdot48\cdot72 \cdot 16$ lattice the smallest eigenvalue of a caloron with maximal nontrivial holonomy and zero chemical potential
reach zero within double precision.

As the staggered operator and its eigenmodes cannot be made real as in the continuum, the argument of the reality of the zero mode profile needs to be revisited. 
From the computations it will turn out that the staggered quasi-zero modes have real parts vanishing to machine precision, such that the quartets practically 
consist of two degenerate pairs, $(\l,-\l^*)$ and $(-\l,\l^*)$, on the imaginary axis. If an eigenvalue $\l$ is degenerate with $-\l^*$ the corresponding subspace is spanned by 
\begin{align}
 \Big(\ps(x;\mu),\eta_5\tau_2\ps^*(x;\mu)\Big)\equiv \Big(\ps_1(x;\mu),\ps_2(x;\mu)\Big)
\label{eq_def_two_deg_modes}
\end{align}
In Appendix \ref{sec:BiOrtho} we show that a particular linear combination of these modes obeys the bi-orthonormalization
\begin{align}
 \int \!\!d^4 x\,\ps_M^\dagger(x;-\mu)\ps_N(x;\mu)=\delta_{MN},\: M,N\in\{1,2\}
\end{align}
and that the profile averaged over these two modes
\begin{align}
  \ch_{00}(x;\mu)=\frac{1}{2}\sum_{M=1}^2\ps_M^\dagger(x;-\mu)\ps_M(x;\mu)
\label{eq_both_modes_density}
\end{align}
is real. Consequently, we will compare the real profiles \eq{eq_both_modes_density} to those in the continuum.

\subsection{Results for eigenvalue spectra}

\begin{figure}[t] 
\includegraphics[width=0.95\columnwidth]{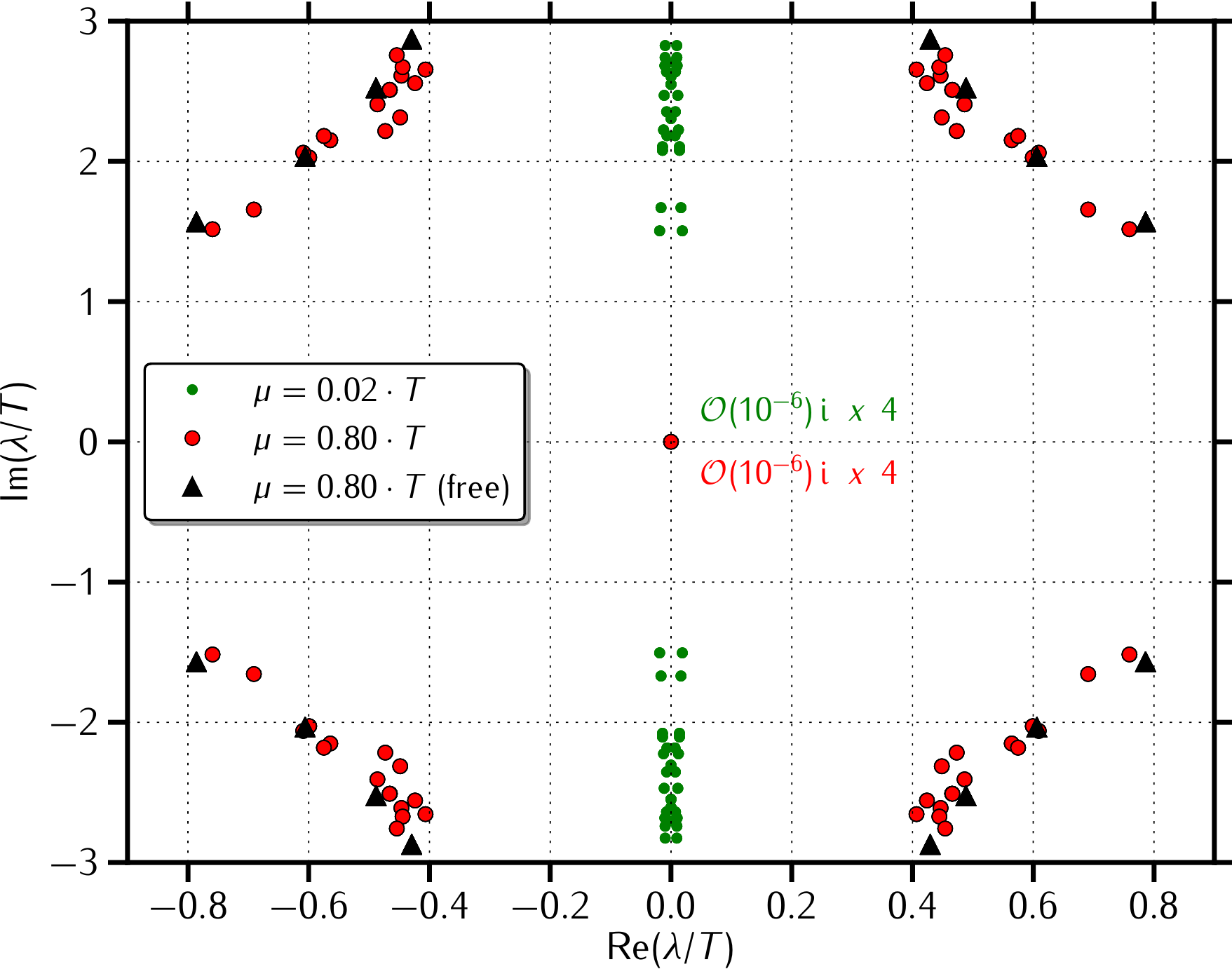} 
\caption{Spectrum of the staggered Dirac operator in a caloron background of maximally nontrivial holonomy ($\omega=0.25$, $\rho= 0.75\beta$)
and different chemical potentials on a $24\cdot24 \cdot 36 \cdot 8$ lattice. 
Note that in order to be able to compute more eigenvalues, this lattice is smaller than the one used for the profiles later in \protect\fig{fig:numzeromode1}.  
For comparison the triangles show the eigenvalues of a free Polyakov loop background with maximal nontrivial holonomy (see text). 
These are highly degenerate.}
\label{fig:spectrum}
\end{figure}

On the configurations described in Sec.~\ref{subsec_discretization} we measured 256 eigenmodes with smallest magnitude by virtue of ARPACK \cite{Lehoucq1997}, Armadillo \cite{Sanderson2010} and the C++ Boost Libraries. 
As expected the eigenvalues of the staggered operator become complex when switching on the chemical potential, shown in \fig{fig:spectrum}. The quasi-zero eigenvalues, however, remain basically unchanged at finite chemical potential. From that figure one further recognizes, that the other low-lying eigenvalues of the caloron are similar to generalized Matsubara frequencies. By that we mean the Dirac spectrum in free backgrounds with constant Polyakov loops of the same holonomy as the caloron
\begin{align}
 U_t=\exp\left(2\pi i \,\text{diag}(\omega,-\omega)/N_t\right)
\end{align}
and trivial space-like links, $U_x=U_y=U_z=\one$,  such that the action vanishes. This background gives rise to the following staggered eigenvalues
\begin{align}\label{free_spectra}
 &\lambda(\mu)/T=\pm\, i \,N_t\times\\ &\sqrt{\sum_{i=x,y,z}\sin^2\left(\frac{2\pi}{N_i}\,k_i\right)+
 \sin^2\left(\frac{2\pi}{N_t}(k_t\pm\omega+\frac{\p-i\mu/T}{2\pi})\right)}\nonumber
\end{align}
with $k_\mu\in(-N_\mu/2,\ldots, N_\mu/2]$ the wave numbers of the plane wave eigenmodes, for vanishing $\mu$ cmp.\ \cite{Bruckmann:2008xr}. For low-lying modes, $|k_\mu|\ll N_\mu$, the sines disappear and the eigenvalues follow the continuum dispersion relation.
 
The similarity of the caloron spectra to these generalized Matsubara frequencies\footnote{A similar similarity holds for the eigenvalues of the gauge covariant Laplacian \cite{Bruckmann:2005hy}.} holds incl.\ nonzero chemical potential, cf.\ \fig{fig:spectrum}. The caloron's quasi-zero modes, which are of topological origin, are of course not reflected in the Matsubara frequencies. 

An interesting effect occurs in the low-lying spectrum when the boundary condition phase $\p$ equals the holonomy $\omega$, such that in the free case they compensate in one color sector (this is the case where the caloron's zero mode cannot be analytically continued since at this boundary condition it hops between the constituents). At vanishing chemical potential the free spectra then possess a zero mode for $\vec{k}=0=k_t$. With nonzero $\mu$, the eigenvalue with these momenta becomes
\begin{align}
 \frac{\lambda(\mu)}{T}=\pm\,N_t\sinh\left(\frac{\mu/T}{N_t}\right)\approx \pm \frac{\mu}{T}
\label{eq:the_real_mode}
\end{align}
and hence is located on the real axis.
With the lowest frequency $k_t=0$, and nonzero $\vec k$, there is a tower of states, which at $\mu=0$ form the low-lying modes on the imaginary axis, while at nonzero $\mu$ they become
\begin{align}
 \frac{\lambda(\mu)}{T}=\pm N_t\sqrt{\sinh^2\left(\frac{\mu/T}{N_t}\right)-\sum_{i}\sin^2\left(\frac{2\pi}{N_i}\,k_i\right)}
\end{align}
In other words, these modes stay on the real axis with magnitudes smaller than the one above and then become purely imaginary with growing magnitudes. How many of these either real or purely imaginary eigenvalues occur on the lattice depends on the aspect ratios $N_i/N_t$. Only when moving to the next frequency $|k_t|=1$ or to the color sector, where $\omega$ does not compensate but adds to the boundary condition, do these eigenvalues become fully complex.  
%\enlargethispage{2\baselineskip}

\begin{figure}[t] 
\includegraphics[width=0.95\columnwidth]{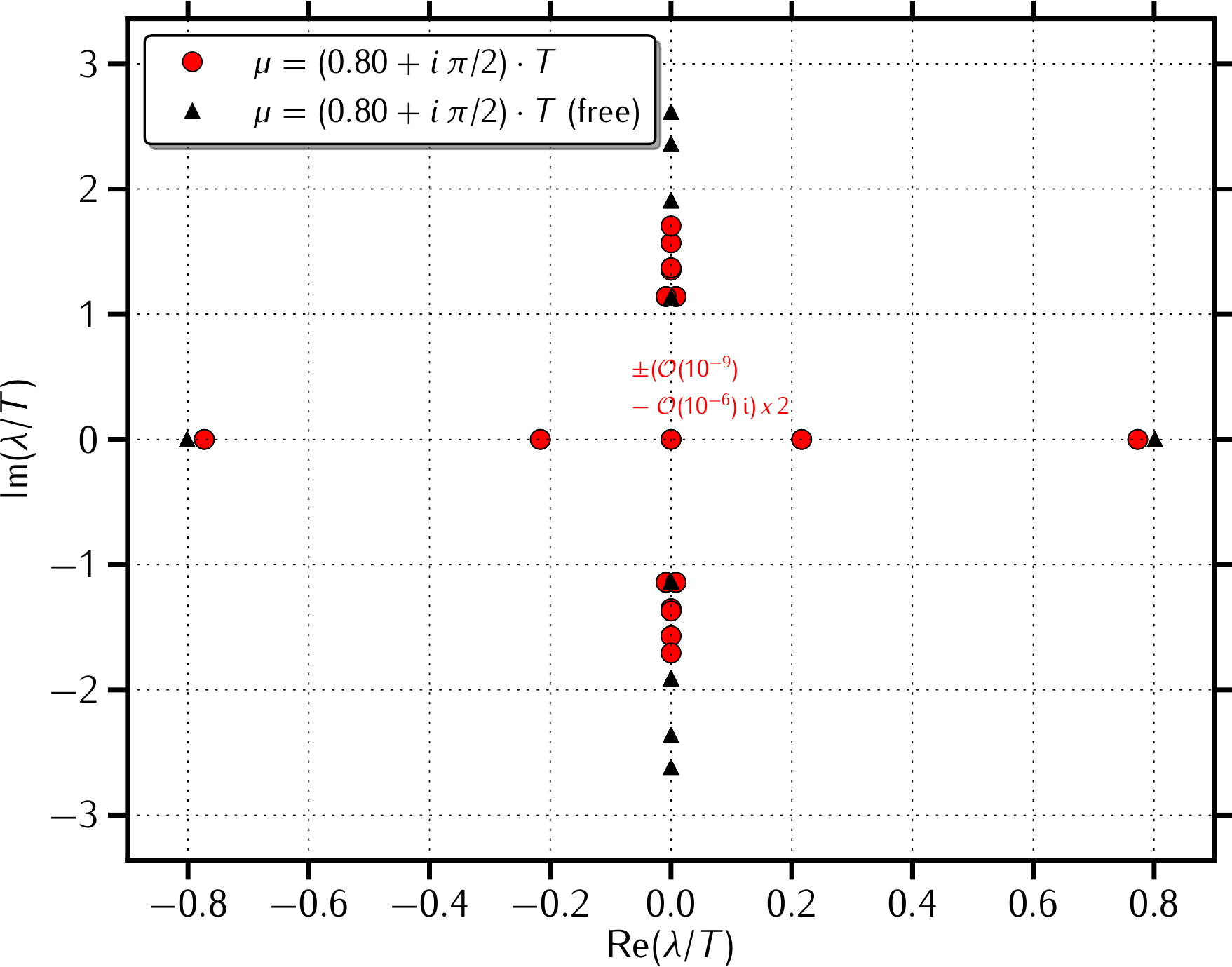} 
\caption{Spectrum of the staggered Dirac operator in a caloron background of maximally nontrivial holonomy (same as in Fig.~\ref{fig:spectrum}) and intermediate boundary condition (periodicity up to $i$) plus  the corresponding free spectrum. The lowest eigenvalues are either real or purely imaginary, see text.}
\label{fig:realorimag}
\end{figure}

The lowest staggered eigenvalues of the caloron in this case are also either real or purely imaginary, see \fig{fig:realorimag} for maximally nontrivial holonomy and boundary phase $\p=\pi/2$ (i.e.\ periodicity up to a factor $i$). One of the caloron's real eigenvalues is again close to its free counterpart (\ref{eq:the_real_mode}).

\subsection{Comparison with the exact zero mode for the caloron}

\begin{figure}
 \includegraphics[width=\columnwidth]{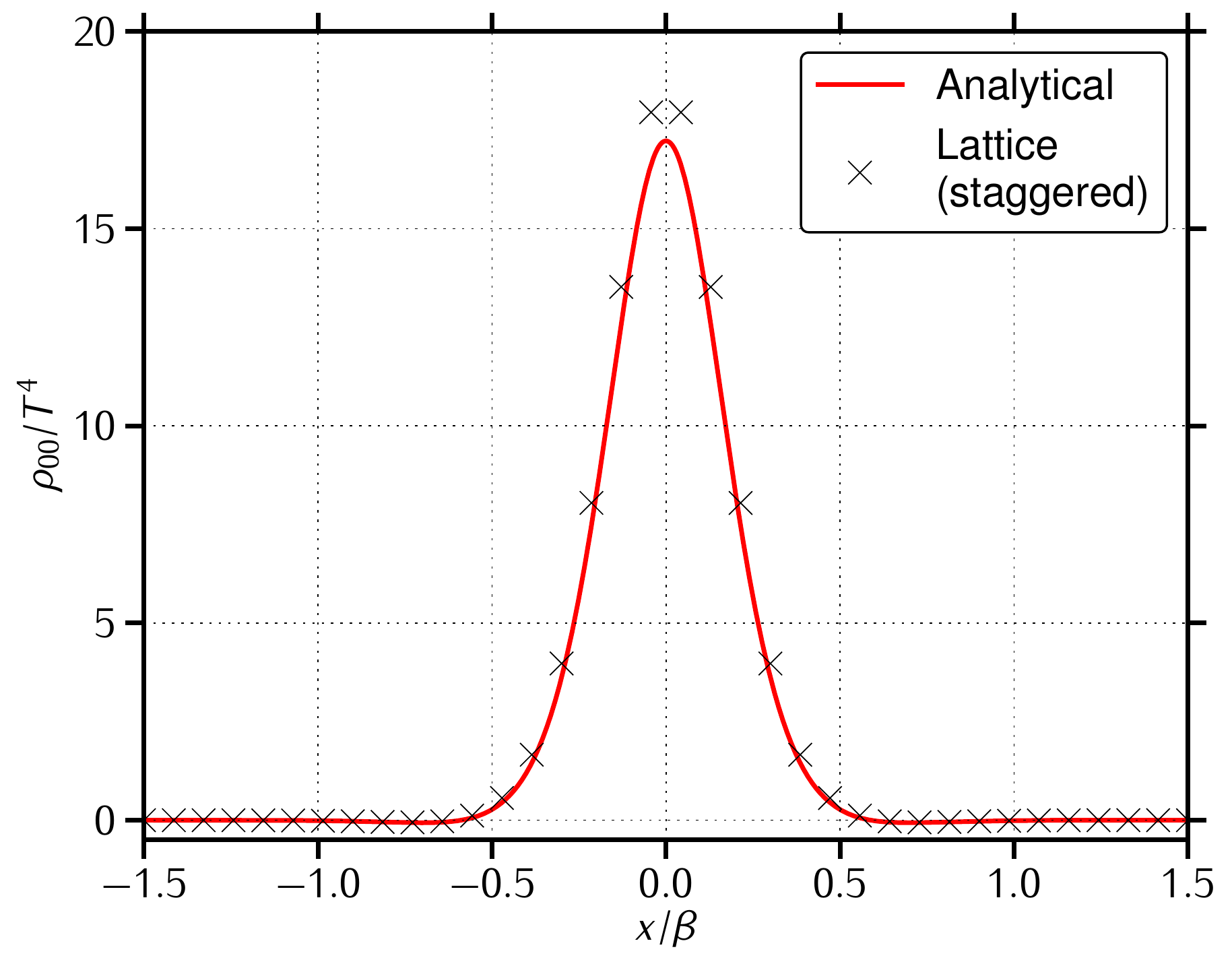}
 \includegraphics[width=\columnwidth]{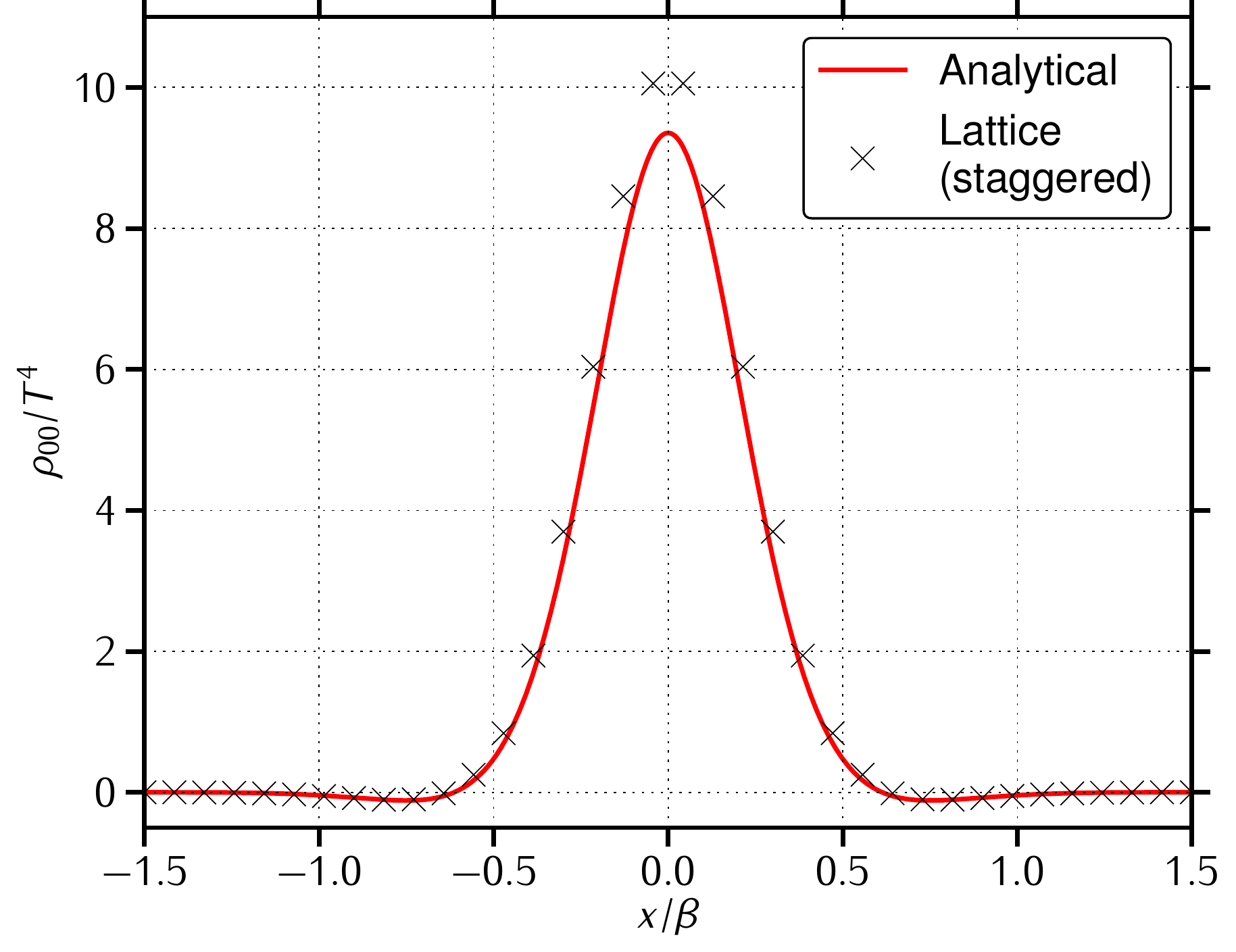}
 \includegraphics[width=\columnwidth]{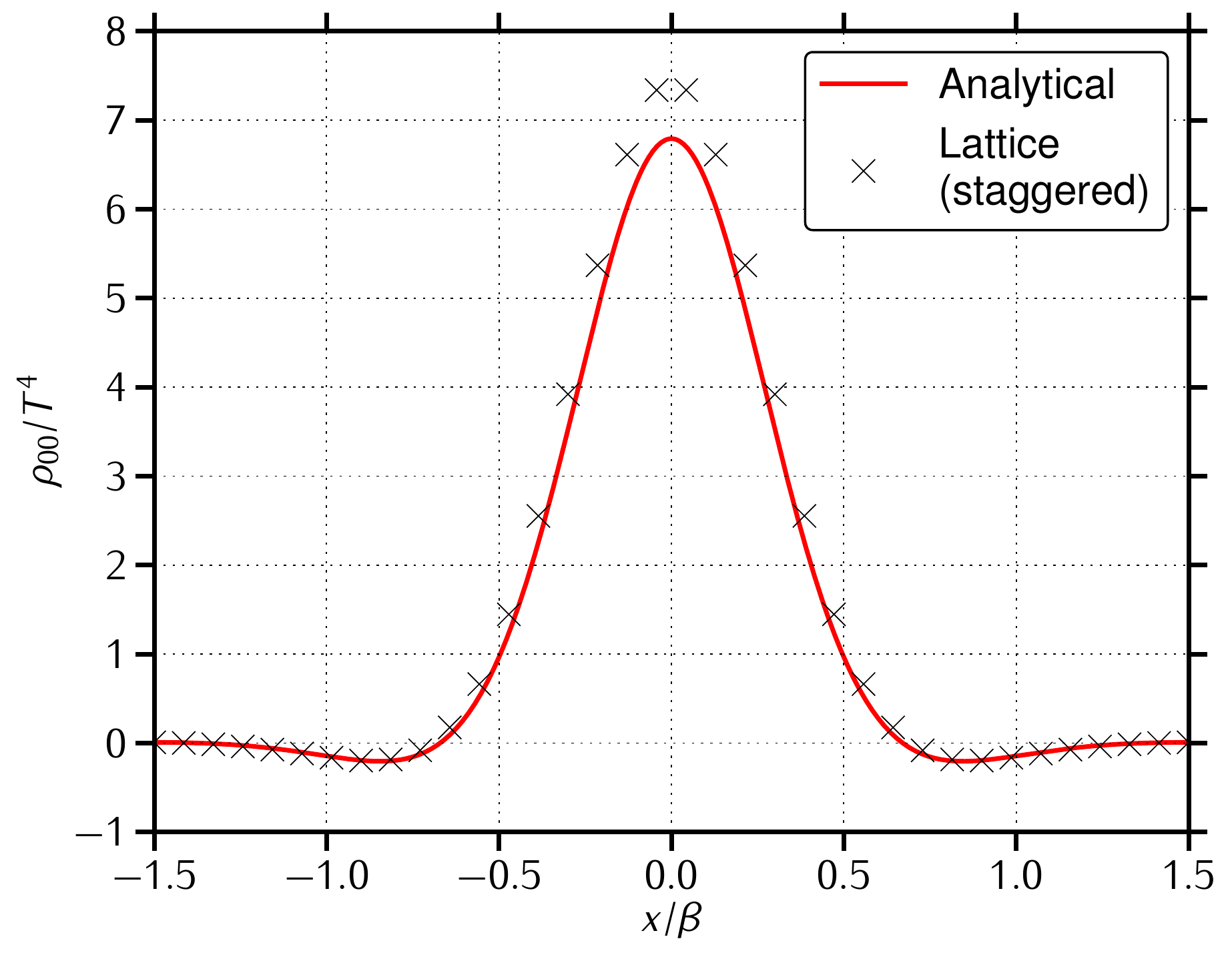}
\caption{Comparison of the analytical and numerical density of the caloron zero mode at $\mu=2.4\,T$, on a line perpendicular to the caloron axis and intersecting with it near the twisted dyon (for more details on the chosen coordinates see text). 
All panels correspond to caloron with size parameter $\rho=0.75\beta$. The holonomy increases from top to bottom: $\omega = 0.0, \, 0.125, \, 0.25$. 
The black crosses are from numerical solutions on a $36\cdot36\cdot54\cdot12$ lattice, while the red solid line correspond to the analytical solution in Eq.~(\ref{eq:density00}).}
%The remaining coordinates are fixed to $t=\beta/(2N_t)$, $y=\beta/(2N_t)$ and $z=0$ (see text). }
\label{fig:numzeromode1}
\end{figure}

Having found quasi-zero modes in the staggered spectrum, we expect the corresponding eigenmodes
to asymptote to the continuum zero modes of \sec{sec_profiles_caloron}.
To get the eigenmodes to be bi-orthonormal, we have computed eigenmodes at $\mu$ and $-\mu$ and used the linear combinations defined in Eq. \eq{eq_both_modes_density}. 
The comparison is shown in \fig{fig:numzeromode1} for fixed chemical potential, $\mu = 2.4 T$, while the holonomy increases from trivial to maximally nontrivial. %: $\omega =  0.0, \, 0.125, \, 0.25$.
To compare the numerical solution to the continuum one, 
we fix the time $t$ and the coordinates $z$ (along the caloron axis) and $y$ (perpendicular to the caloron axis) as close as possible to the twisted dyon's core, on which the (antiperiodic) zero mode is localized. Our lattices are shifted such, that the caloron axis runs through the middle of a plaquette. Therefore, we choose $y=t=a/2=\beta /(2N_t)$. In order to show two-dimensional plots, we finally fix the $z$-coordinate. We choose that value of $z$, for which the numerical density $\ch_{00}(x;\mu)$ (cf.\ \eqref{eq_both_modes_density}) as a function of the remaining coordinate $x$ is maximal. The analytic density (\ref{eq:density00}, \ref{eq:ffunctwithmu})) is plotted along the corresponding continuum line. For holonomies $\omega = 0.0, \, 0.125, \, 0.25$ this amounts to shifts in $z$ of $0.5a$, $0.8a$ and $1.1a$ from the formal dyon core $r=0$ towards the other dyon.

We find that the numerical densities coincide well with the continuum zero mode densities.
In particular, the negative regions are confirmed by the numerical solutions ($\omega = 0.125, \, 0.25$ in \fig{fig:numzeromode1}). 
That for increasing holonomy the mismatch becomes stronger can be explained by considering the action densities of those calorons.
For $\omega = 0$ there is just one dyon which is far away from the boundaries of the lattice, 
while in the case of maximal nontrivial holonomy two separated dyons
are in the same box. This means that the finite volume effects are stronger in the case of maximally nontrivial holonomy. 

\section{Overlap matrix elements}
\label{sec_overlap}

A very important quantity for building a dyon liquid model in analogy to the instanton liquid model is the overlap matrix element \cite{Schafer:1998up,Cristoforetti:2011fq}. 
\be
T_{I\bar J}=\int d^4x\;\psi^\dagger_{I}(x;-\mu)(-i\slashed\partial+i\mu\gamma_0)U\psi_{\bar J}(x;\mu)\;,
\ee
where $\psi_{I},\psi_{\bar J}$ are zero modes of dyons labeled by the index $I$ and of antidyons labeled by the index $\bar J$, respectively. This form reflects the residual $U(1)$ gauge freedom, as the two dyons can always be superposed in gauges differing by a quasi-abelian gauge transformation $U=\exp(i\alpha\hat\omega\cdot\vec\tau)$ which preserves the holonomy. This replaces the general $SU(2)$ relative gauge transformation in the case of the instanton anti-instanton overlap matrix element (see \cite{Schafer:1998up}).

\begin{figure*}[!t] 
   \centering
  \includegraphics[width=3.5in]{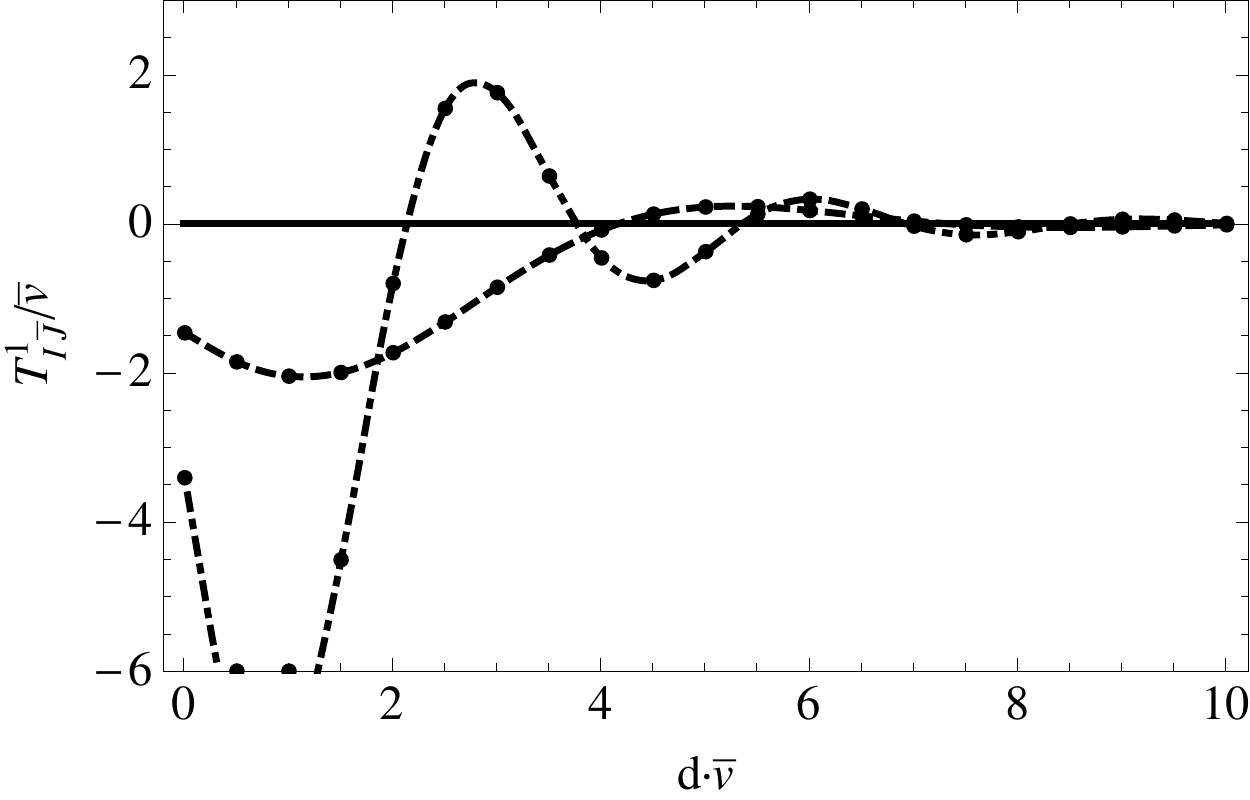}    
  \includegraphics[width=3.5in]{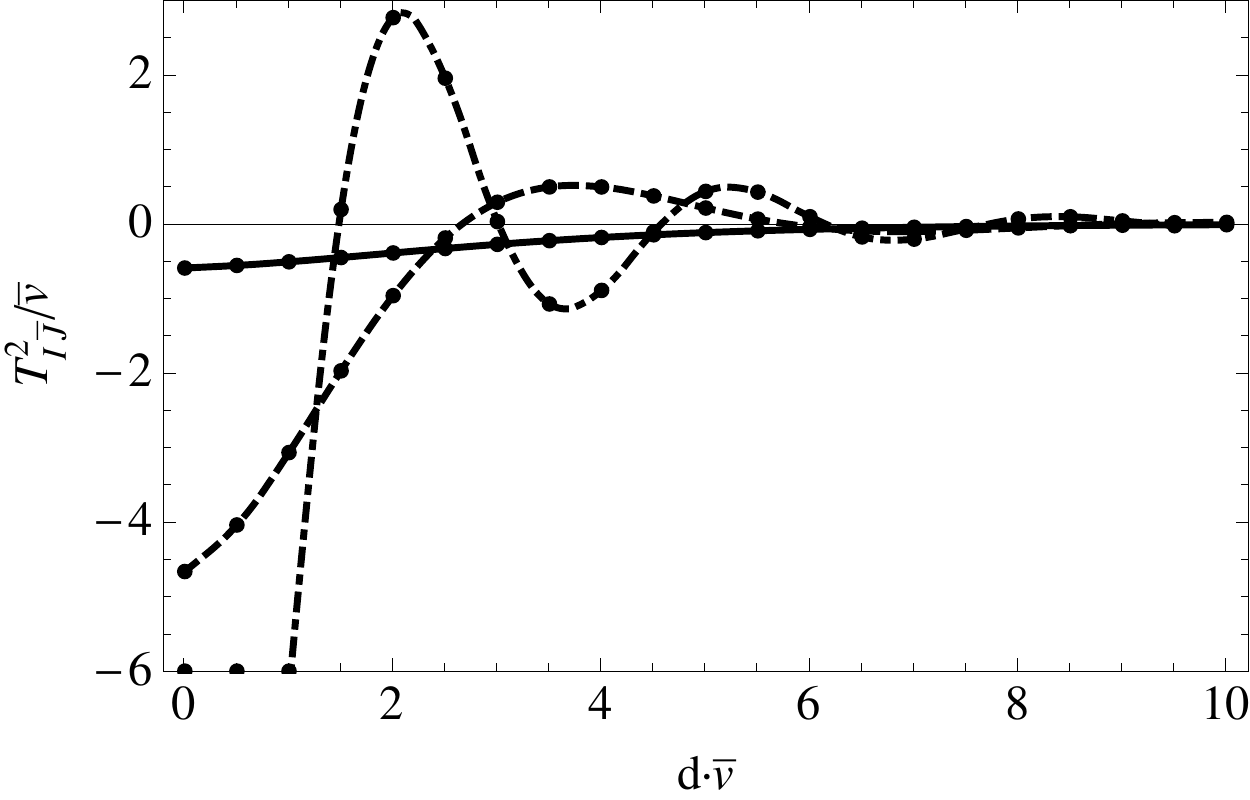}  
   \caption{Plots of numerical integration of functions $T_{I\bar J}^{1,2}$, \eqref{eq:T_decomp}, as a function of distance $d$ in units of $\bar v$. The plots show results for $\mu=0$ (solid), $\mu=\bar v$ (dashed) and $\mu=2\bar v$ (dot-dashed). Note that the behavior is similar to that of the behavior of two spatially separated instantons (see Fig. 1 in \cite{Schafer:1998up})}  
   \label{fig:hoppingnum}
\end{figure*}

\begin{figure}[b]
\includegraphics[width=.4\textwidth]{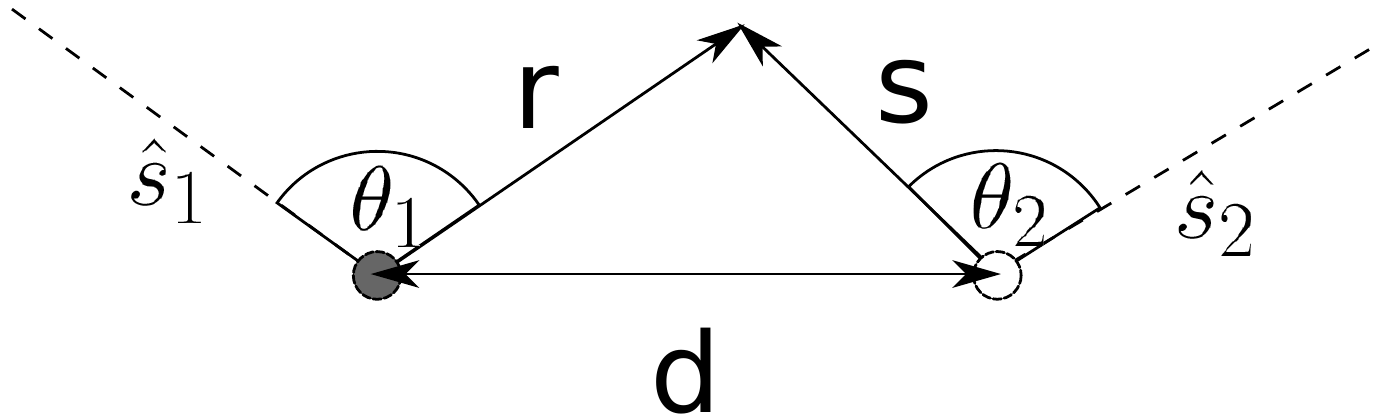}
\caption{Configuration of dyon (shaded) and antidyon (white) and their Dirac strings (dashed lines) discussed in the text.}
\label{fig:hopping1}
\end{figure}

Our formulas \eqref{eq:componentform} apply only for the case that holonomy is chosen along the line that connects two constituent dyons, and that this line is chosen along the $z$-direction. This also applies to the single dyon limit of the previous subsection. It is easy to generalize this expression to the arbitrary direction of the Dirac string
\begin{align}
&\psi_I(t,\vec{r};\mu)=\frac{\sqrt{T}}{2\sqrt{2\pi r}}\frac{e^{\mu t}}{\sqrt{\coth(4\pi r\bar\omega T)-\cos\theta_1}}
 \nonumber\\
&\hspace{.5cm}\times\partial_\nu\Big\{e^{-\mu t}\Big(\frac{\cos(r\mu)}{\cosh(2\pi r \bar\omega T)}\nonumber\\
&\hspace{.5cm}-i\hat{s}_1\cdot\vec{\tau}\frac{\sin(r\mu)}{\sinh(2\pi r \bar\omega T)}\Big)e^{2\pi i t \bar\omega\hat{s}_1\cdot\vec{\tau}T}\bar\sigma_\nu\epsilon\Big\}\;,
\end{align}
\begin{align}
&\psi_{\bar J}(t,\vec{s};\mu)=\frac{\sqrt{T}}{2\sqrt{2\pi r}}\frac{e^{\mu t}}{\sqrt{\coth(4\pi s\bar\omega T)-\cos\theta_2}}
 \nonumber\\
&\hspace{.5cm}\times\partial_\nu\Big\{e^{-\mu t}\Big(\frac{\cos(s\mu)}{\cosh(2\pi s \bar\omega T)}\nonumber\\
&\hspace{.5cm}+i\hat{s}_2\cdot\vec{\tau}\frac{\sin(s\mu)}{\sinh(2\pi s \bar\omega T)}\Big)e^{-2\pi i t \bar\omega\hat{s}_2\cdot\vec{\tau}T}\sigma_\nu\epsilon\Big\}\;.
\end{align}
where now $\theta_{1,2}$ are the angles between $\vec{r},\vec{s}$ and the Dirac strings\footnote{Note that because for a selfdual field $A_\mu(t,\vec{x})$, we have that $\tilde A_\mu(t,\vec{x})=(A_0(t,-\vec{x}),-\vec{A}(t,-\vec{x}))$ is anti-selfdual. That means that if $\slashed D(A)\psi=0$ then $\slashed D(\tilde A)\gamma_0\psi(t,-\vec{x})=0$, so one can construct an antidyon (anti-caloron) zero mode solution from the dyon (caloron) zero mode solution by flipping $\vec{x}\rightarrow -\vec{x}$ and multiplying by $\gamma_0$. Note that when we construct the antidyon zero mode in this way we also have to flip the direction of the Dirac string.} ${\hat s}_{1,2}$ respectively (see Fig. \ref{fig:hopping1}). Note that we superpose in the algebraic gauge in which $A_0\to 0$ away from the dyons, cf.\ \sec{sec:cal}. The dyons in question are those that carry the anti-periodic zero modes (the ``heavy'' ones), although the same procedure can easily be adapted to the case of dyons carrying the periodic zero modes.

However, the above solutions assume holonomies in the directions $\hat{s}_{1}$ and $-\hat{s}_2$ for the dyon and antidyon. To compensate that we define rotation matrices 
\begin{align}
&\mathcal U_1 \hat{s}_1\cdot\vec{\tau}\, \mathcal U_1^\dagger=\hat\omega\cdot\vec{\tau}\;,\\
&\mathcal U_2 \hat{s}_2\cdot\vec{\tau}\, \mathcal U_2^\dagger=-\hat\omega\cdot\vec{\tau}\;,
\end{align}
where $\hat\omega$ is an arbitrary unit vector parametrizing the final direction of the Polyakov loop in color space (which without loss of generality can be chosen in the third direction). So what we need to compute for the overlap matrix element is
\be
T_{I\bar J}=\int d^4x\; \psi_I^\dagger (x;-\mu)\,\mathcal U_2^\dagger U\mathcal U_1(-i\slashed\partial+i\mu\gamma_0)\psi_{\bar J}(x;\mu)
\ee
where $U=\exp(i\alpha \hat\omega\cdot\vec{\tau})$ is the relative quasi-abelian gauge transformation between the dyons mentioned already at the beginning of this section. Note that for instanton a relative $SU(2)$ gauge transformation was relevant. Here the relative quasi-abelian gauge transformation is relevant, while the rest of the subgroup $SU(2)/U(1)$ reduces to the orientation of the strings (i.e.\ angles between strings and holonomy).

\begin{figure*}[!t] 
   \centering
   \includegraphics[width=3.5in]{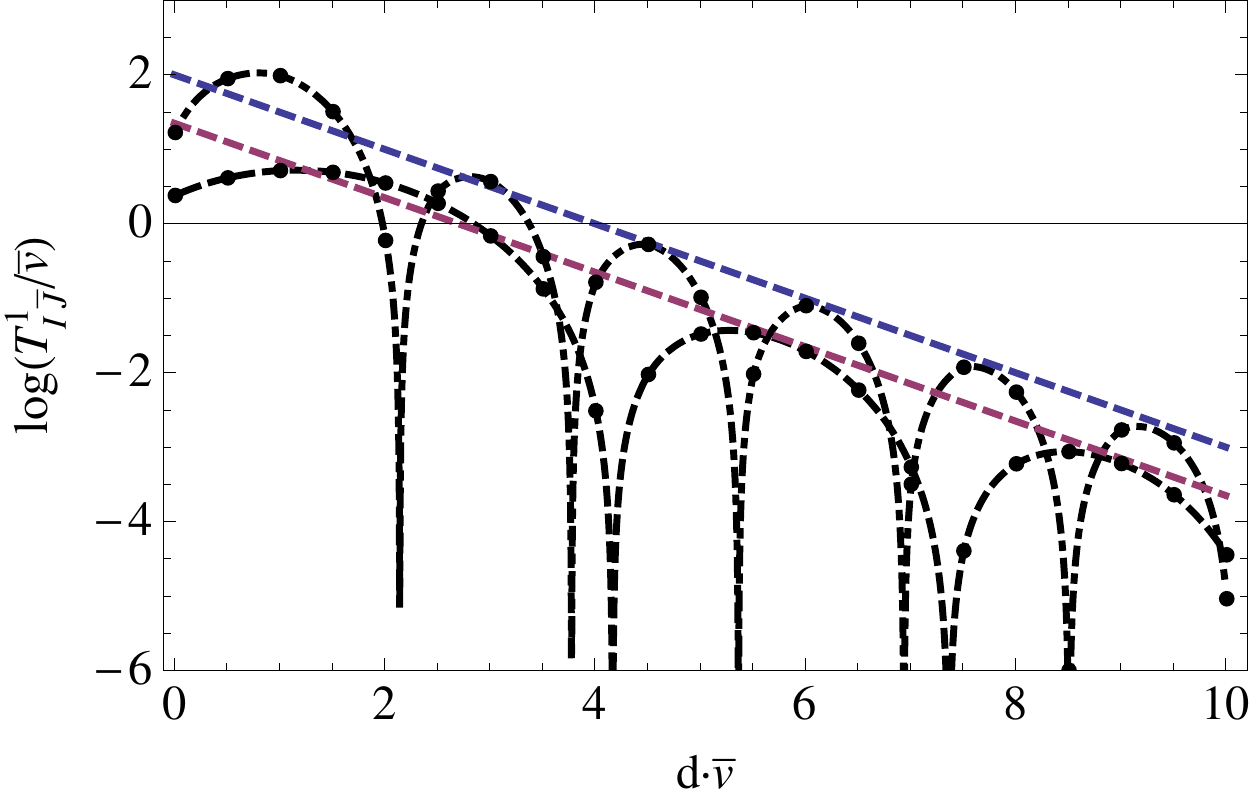}    \includegraphics[width=3.5in]{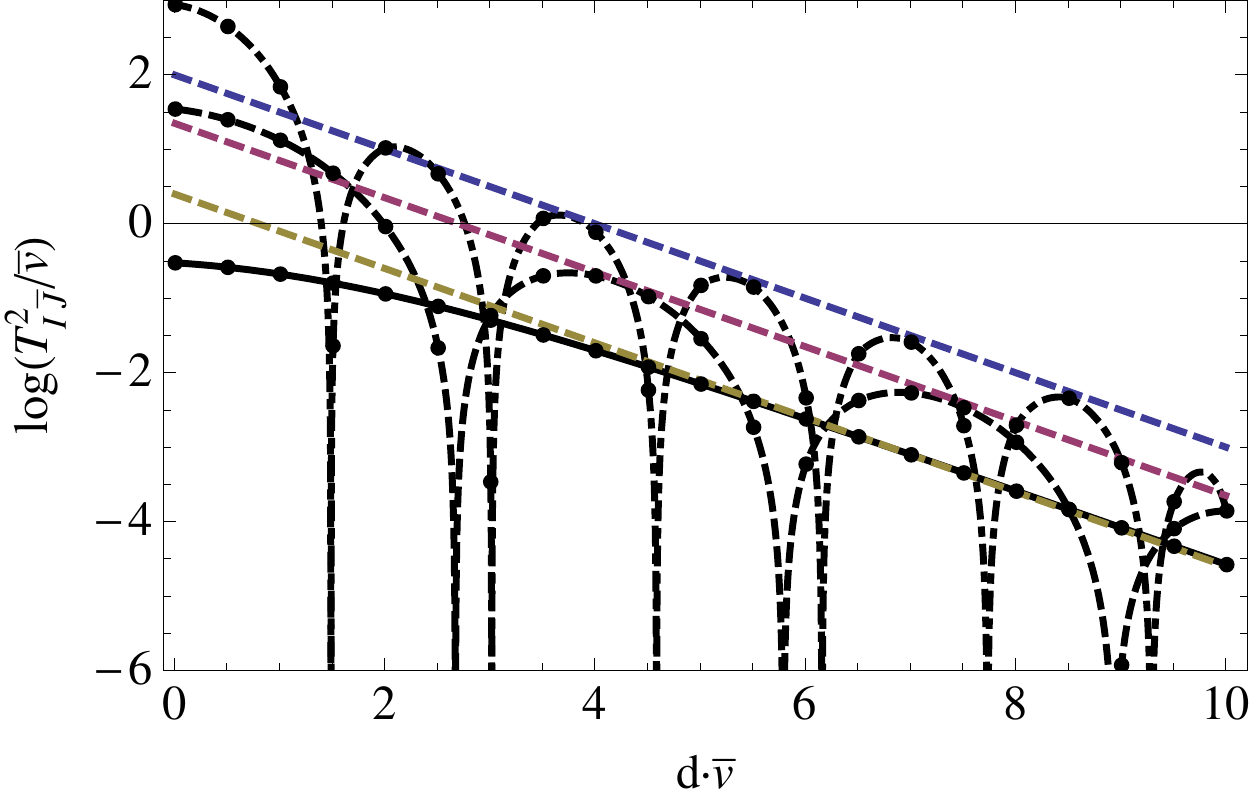} 
   \caption{Logarithmic plots of $T_{I\bar J}^{1,2}$ as a function of distance $d$ in units of $\bar v$, for $\mu=0$ (solid) $\mu=\bar v$ (dashed) and $\mu=2\bar v$ (dot-dashed).  The dashed, straight lines are linear functions with slope $-1/2$ to guide the eye.}
   \label{fig:hoppingnumlog}
\end{figure*}

We leave the problem of the hopping element of a caloron in full generality for future work, but here to illustrate the qualitative behaviour let us take a simple configuration of the dyon-antidyon system with their strings lying on the $z$-axis, and pointing away from each other.

Writing the overlap matrix element 
\begin{align}
 T_{I\bar J}=i(T_{I\bar J}^1\cos\alpha+T_{I\bar J}^2\sin\alpha)
\label{eq:T_decomp}
\end{align}
it is not very difficult to see that $T_{I\bar J}^1=0$, at $\mu=0$. This happens simply because of the index structure of the above expressions, resulting, upon index contraction, in vanishing contribution. For $\mu\ne0$ this is no longer the case, as there is an additional term proportional to $\tau_3$ in the expression for the zero mode (see \eqref{eq:psi0_single}).  In Fig. \ref{fig:hoppingnum} and \ref{fig:hoppingnumlog} we show the results of numerical integration of the overlap matrix element as a function of distance between the dyon and anti-dyon.

Because of exponential localizations, the most dominant contribution to the overlap matrix at large distance would naively come from the two regions where the zero modes are localised. However since they decay exponentially, along the line connecting the dyon and antidyon, there is a significant contribution in this region, turning the asymptotic behavior into exponential $\sim\exp(-\bar v d/2)$ (with no algebraic factors).

\section{Summary}
\label{sec_summary}

We have analyzed zero modes at nonzero chemical potential in topological backgrounds. Starting with nontrivial holonomy calorons, for which we have analytically continued known zero modes from imaginary chemical potential (i.e.\ phase boundary conditions in the temporal direction), we have derived the zero modes for trivial holonomy calorons and instantons known in the literature as well as for dyons, for which a spherical ansatz was explored, too. All of these procedures have a direct generalization from SU(2) considered here to SU(N). Zero modes present at $\mu=0$ due to index theorems remain at nonzero $\mu$. 

The most prominent features of these zero modes are higher values at their centers and negative regions in their (pseudo-)density. In dyon backgrounds the former increases asymptotically like $\mu^4$, while the distance between the zeros between regions of different sign is  proportional to $1/\mu$. The envelop of the decay is to leading order independent of $\mu$. 

On lattice-discretized calorons we have found completely analogous quasi-zero modes for the staggered Dirac operator. For the latter this means that although exact chiral symmetry is lacking, the topological part of the continuous spectrum on such smooth backgrounds is well-reflected (in eigenmode quartets) also at nonzero chemical potentials. Therefore, this numerically cheap operator could be sufficient for the analysis of ensembles of topological ensembles (or realistic lattice Monte Carlo configurations) at nonzero density.

The overlap matrix elements of such zero modes are a key ingredient to the interaction of such topological objects. As we have demonstrated by virtue of numerical integration, the exponential decay of the overlap matrix elements with the distance are the same as for zero $\mu$, but their alternating behavior will cause an additional wash-out effect similar to that of instantons \cite{Schafer:1998up}. We also showed that a term in the overlap matrix element appears, that is related to the relative abelian orientation of the dyons and that vanished at $\mu=0$. More consequences of the overlap matrix elements at nonzero chemical potential are left for future studies.  

\section{Acknowledgements}

This work has been supported by DFG (BR 2872/4-2), the Alexander von Humboldt foundation and a scholarship of BayEFG (T.S.). 

\begin{appendix}

\section{Reality of profiles in the continuum}
\label{sec:reality}

If $\ds(\mu)$ vanishes on the zero mode $\ps_0(\mu)$, then from the conjugate equation and the anti-unitary SU(2) symmetry (\ref{anti_symm_cont}) it follows that $\ds(\mu)$ vanishes on $W\ps_0(\mu)^*$ as well. For a nondegenerate zero mode the latter must be proportional to the zero mode up to a factor
\begin{align}
 W\ps_0(\mu)^*=\al(\mu)\ps_0(\mu)
\label{eq:relation_modes}
\end{align}
This factor $\al$ is indeed a phase, which can be seen by acting with $W$ on the complex conjugate equation using $WW^*=1$ and the previous equation again:
\begin{align}
 W W^*\ps_0(\mu)&=W \al(\mu)^*\ps_0(\mu)^*\nonumber\\
 \ps_0(\mu)&=\al(\mu)^*\al(\mu)\ps_0(\mu)
\end{align}
Next we prove that $\al(-\mu)=\al(\mu)$. For that we use the bi-orthonormalization at $\pm\mu$,  $W^T=W$ and (\ref{eq:relation_modes}) at $-\mu$ to write
\begin{align}
 \alpha(\mu)&=\int \!d^4x\, \ps_0^\dagger(-\mu)W\ps_0(\mu)^*\nonumber\\
 \alpha(\mu)^T&=\int \!d^4x\,\ps_0^\dagger(\mu)W^T\ps_0(-\mu)^*\nonumber\\
 \alpha(\mu)&=\int \!d^4x\,\ps_0^\dagger(\mu)\al(-\mu)\ps_0(-\mu)=\alpha(-\mu)
\end{align}
Finally, from (\ref{eq:relation_modes}), its hermitian conjugate and the unitarity of $W$:
\begin{align}
\Big[\ps_0^\dagger(-\mu)\ps_0(\mu)\Big]^*
 &=\ps_0^\dagger(-\mu)\al(-\mu)^* W W^\dagger\al(\mu)\ps_0(\mu)\nonumber\\
 &=\ps_0^\dagger(-\mu)\ps_0(\mu)
\end{align}
which just means that the density $\rho_{00}(\mu)$ is real for all $\mu$ and $x$.

\section{Bi-orthonormalization and reality for degenerate staggered eigenmodes}
\label{sec:BiOrtho}

With the definition \eq{eq_def_two_deg_modes} the densities in the bi-orthonormalization 
\begin{align}
 \ps_M^\dagger(x;-\mu)\ps_N(x;\mu)\equiv\ch_{MN}(x;\mu)
\end{align}
obey
\begin{align}
 \ch^\dagger(x;\mu)=\ch(x;-\mu)
 \label{eq_ch_dagger}
\end{align}
and have the form of a quaternion
\begin{align}
 \ch(x;\mu)
  &\equiv\left(\begin{array}{cc}g(x;\mu)&h(x;\mu)\\-h^*(x;\mu)&g^*(x;\mu)\end{array}\right)\\
 \int \!\!d^4 x\,\ch(x;\mu)
  &\equiv\left(\begin{array}{cc}G(\mu)&H(\mu)\\-H^*(\mu)&G^*(\mu)\end{array}\right)
\end{align}
i.e.\ the modes are not necessarily bi-orthogonal (unless $H=0$).

Choosing two different modes in the subspace amounts to a right multiplication
\begin{align}
 \ps_M(x;\mu)\to\ps_N(x;\mu)C_{NM}(\mu)\quad C(\mu)\in GL(2,C)
\end{align}
where the coefficient matrix $C(\mu)$ may depend on $\mu$, but must be a constant wrt. $x$. The density transforms as
\begin{align}
 \ch(x;\mu)\to C^\dagger(-\mu)\ch(x;\mu)C(\mu)
\end{align}
preserving the property \eq{eq_ch_dagger}. The integral transforms accordingly,
\begin{align}
 \int \!\!d^4 x\,\ch(x;\mu)\to C^\dagger(-\mu)\int \!\!d^4 x\,\ch(x;\mu)\,C(\mu)
\end{align}
With the choice
\begin{align}
 C(\mu)=\Big[\int \!\!d^4 x\,\ch(x;\mu)\Big]^{-1/2}=C^\dagger(-\mu)
\end{align}
one achieves the desired bi-orthonormalization
\begin{align}
 \int \!\!d^4 x\,\ch(x;\mu)\to 1_2
\end{align}
and for the sum of the densities
\begin{align}
 \sum_{M=1}^2&\ps_M^\dagger(x;-\mu)\ps_M(x;\mu)
  \to \tr\, C^\dagger(-\mu)\ch(x;\mu)C(\mu)\nonumber\\
  =\,&\tr\,\left(\begin{array}{cc}G&H\\-H^*&G^*\end{array}\right)^{-\frac{1}{2}}
   \left(\begin{array}{cc}g&h\\-h^*&g^*\end{array}\right)
   \left(\begin{array}{cc}G&H\\-H^*&G^*\end{array}\right)^{-\frac{1}{2}}\nonumber\\
  =\,&\frac{Gg^*+Hh^*}{|G|^2+|H|^2}+c.c.
\end{align}
which is real. That this expression is the sum of two densities is accounted for by a factor $1/2$ in \eqref{eq_both_modes_density}.

\end{appendix}

\bibliography{topmu}
\bibliographystyle{h-physrev}

\end{document}